\documentclass[aps,prb,reprint,amsmath,amssymb,superscriptaddress,longbibliography,onecolumn]{revtex4-2}
\usepackage{graphicx}
\usepackage{mathrsfs}
\usepackage{bm}
\usepackage{textcomp,color}
\usepackage[colorlinks=true,urlcolor=blue,citecolor=blue,linkcolor=blue,bookmarks=false, pdfstartview={FitH}]{hyperref}
\usepackage[capitalize]{cleveref} % more convenient citations
\usepackage{dsfont}
\usepackage{amsmath}
\usepackage{float}
\usepackage{tabularx}
\usepackage{multirow}
\usepackage{booktabs}
\usepackage{hhline}

\usepackage[normalem]{ulem} % for \sout

\newcommand{\beq}{\begin{equation}}
\newcommand{\eeq}{\end{equation}}
\newcommand{\bea}{\begin{eqnarray}}
\newcommand{\eea}{\end{eqnarray}}
\newcommand{\bse}{\begin{subequations}}
\newcommand{\ese}{\end{subequations}}
\newcommand{\nn}{\nonumber}
\newcommand{\bwt}{\begin{widetext}}
\newcommand{\ewt}{\end{widetext}}

\newcommand{\ve}{\varepsilon}
\newcommand{\e}{\epsilon}

\newcommand{\bk}{{\bf k}}
\newcommand{\bK}{{\bf K}}

\newcommand{\bq}{{\bf q}}
\newcommand{\br}{{\bf r}}
\newcommand{\bR}{{\bf R}}

\begin{document}

\title{A generalized framework for straintronics in 2D quantum materials using group theory}

\author{Rami Zemouri}
\affiliation{Department of Physics, Concordia University, Montréal, QC H4B 1R6, Canada}
\author{A. R. Champagne}
\affiliation{Department of Physics, Concordia University, Montréal, QC H4B 1R6, Canada}
\author{Saurabh Maiti}
\affiliation{Department of Physics, Concordia University, Montréal, QC H4B 1R6, Canada}
\affiliation{Centre for Research on Multiscale Modelling, Concordia University, Montréal, QC H4B 1R6, Canada}
\date{\today}

\begin{abstract}
In the era of 2D and quasi-2D quantum materials one needs to model strain at the level of the Hamiltonian as opposed to a semi-classical approach. Corrections to the electronic Hamiltonian due to strain arise from two sources: deformations of the lattice and changes in the hoppings. Here, we provide a general theory that takes into account the symmetry of the lattice and that of the bonds, and allows us to write down the strain corrections from all sources in any 2D lattice in terms of the band structure parameters like the velocity and the inverse mass tensor. We then use Group theory to identify when strain can be described as a scalar- and/or a vector-potential. We discuss the nature of the potentials that arise from in- and out-of-plane hoppings, allowing us to model multi-layer systems. We also show that, in general, one encounters multiple vector-potentials in different sectors of the Hilbert space, but one can derive a simpler effective low energy strained Hamiltonian via Hilbert space projections. We discuss several toy models of 2D systems and present strained bilayer graphene as a practical system that incorporates all of the above features. We identify a strain dependent energy scale in bilayer graphene above which we would no longer need to account for multiple vector-potentials. The generality of this formulation allows for a wider range of materials to be investigated for quantum transport straintronics.
\end{abstract}

\maketitle
\tableofcontents

\section{Introduction}\label{Sec:Introduction}
Strain is inherent when working with two dimensional (2D) systems, and is unavoidable due to the presence of a substrate, or contacts, or even thermal changes. It can also be deliberately introduced to control/tune the system's behavior.  Some examples of this are the strain induced zero field quantum Hall effect \cite{Guinea2010}, strain induced spin-accumulation \cite{Kato2004}, photonic spin-Hall effect in strained Weyl semi-metals \cite{Jia2021}, strain induced quantum phase transitions \cite{Parker2021}, strain induced spontaneous time reversal symmetry breaking superconductivity \cite{Roy2014,Grinenko2021}, topological flat bands from periodic strain\cite{Wan2023}, to name a few. An equivalent level of interest has been shared on the applied-physics front with developments in 2D straintronics \cite{Shioya2015,Zhang2019,Wang2020,miao_straintronics_2021,Kogl2023,kim_strain_2023}, where transport characterization of strained graphene devices has garnered interest \cite{McRae2019,wang_global_2021,McRae2024,Huang_2024}. In the context of graphene, in particular, the effect of strain has been studied in relation to phonons \cite{Suzuura2002,Manes2007}, rippled membranes \cite{Kim_2008} and induction of a vector-potential \cite{Pereira2009} leading to a pseudo magnetic field. The space of 2D materials of interest has also been extended to include multiple layers \cite{Parhizkar2022} of graphene, of transition metal di-chalcogenides (TMDCs) \cite{Peng2020}, twisted layers of triangular lattice systems \cite{Kazmierczak2021}, twisted layers of strongly correlated \textit{square} lattice systems like twisted cuprates \cite{Zhao2023}. These are all situations where strain should play an important role. Recently, there has also been an interest to study the role of strain in photonic lattices \cite{Qi2023}.

Despite such a ubiquitous presence, and its ability to significantly modify the electronic properties of quantum materials \cite{kim_strain_2023}, one notes that there isn't a generalized scheme to model the impact of strain. While there are a variety of techniques that people have used to study the effect of strain, they have usually been specific to materials. These techniques include the ab-initio methods \cite{Gui2008,Choi2010} which are numerically intensive and not viable for use in building universal models to understand experimental data. There are tight-binding schemes \cite{ramezani_masir_pseudo_2013,de_juan_gauge_2013,Oliva-Leyva_2017} which are relatively more tractable, but they have mostly focused on graphene. For slow variations, continuum level field theoretical modelling \cite{vozmediano_gauge_2010,Arias2015} can be done which allows for studying time-dependent strain \cite{Vaezi2013,Bhat2018}, but these have also mostly focused on graphene, in particular, the Dirac nature of the Hamiltonian. Thus, as it stands to date, to model strain, one needs to start from scratch should one need to investigate a new system. Our formulation addresses this limitation.

Strain corrections are known to arise from two sources \cite{oliva-leyva_understanding_2013}: the modification of the translation vectors due to deformation of the lattice, and the modification of hopping parameters which results from the modifications to the relative distances between atoms. We shall refer to these as being displacement ($D$) induced and hopping ($H$) induced, respectively. In this work, using a tight-binding perspective, we write down the general strain correction in terms of the band structure parameters of the Hamiltonian like the velocity and the inverse mass tensor. It is shown that the $D$ induced corrections couple universally through the velocity while the $H$ induced terms do so through the inverse mass. Next, using group theory, we classify these strain corrections into potentials belonging to different irreducible representations (irreps) of a ``bond-wavevector group". This is a common subgroup of the point group of a bond and the wavevector group \cite{Dresselhaus2007-xs} at a specific point of interest in the Brillouin zone (BZ). We are able to identify when the potentials serve like a scalar-potential, a vector-potential, or a potential in other irreps of bond-wavevector group. In particular it is shown that a vector-potential like interpretation for in-plane deformations is only possible if the bond-wavevector group is $C_{1v}$ or $C_{3v}$.

\begin{figure}[ht]
    \centering
    \includegraphics[width=\textwidth]{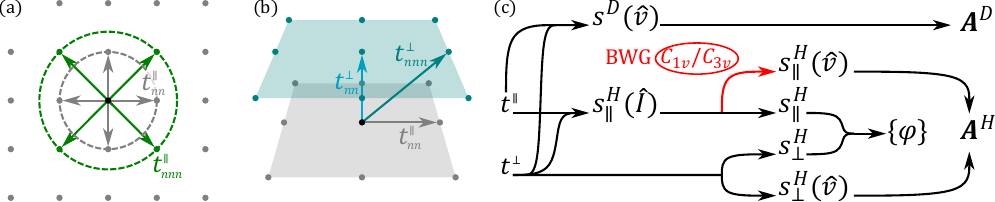}
    \caption{\textbf{Strain induced potentials from hoppings.} Hoppings of different orders [nn, nnn, in($\parallel$)- and out-of-plane($\perp$)] in (a) a 2D lattice,  and (b) a bilayer system. The different orders are marked by different colors. When strained, they induce a potential according to the flow chart in (c). Hoppings of all orders produce potentials in the scalar and vector irreps. The $D$ induced correction universally induces a vector-potential. The in-plane hopping induced correction only produces a vector-potential when the bond-wavevector group (BWG) is $C_{1v}/C_{3v}$, while there is no such restriction on out-of-plane hoppings.} \label{Fig:1}
\end{figure}

In addition to these general results, we show that the vector-potential from non-uniform deformations that are $D$ induced do not have a curl, while that from $H$ induced ones do and hence can induce a pseudo magnetic field. We also explore the role of deforming an out-of-plane bond and show that it can contribute to a vector-potential in the presence of an out-of-plane shear, something that can be realized in hetero-strain scenarios \cite{Hou2024} where the different layers are subject to different strains. We show that each order of hopping (classified based on distance and the type of bonds) induces a set of scalar- and vector-potentials. This is illustrated in Fig. \ref{Fig:1} where panels $(a),(b)$ show examples of bonds of different orders in different colors. In panel $(c)$ we see a chart that outlines how each hopping with in- and out-of-plane hoppings contribute to the respective strain corrections $s^D/s^{H}_{\parallel/\perp}$ and associated potentials $\varphi$ and $\boldsymbol{\mathcal A}$, with the latter being the vector-potential. Our theory thus suggests, quite naturally, the presence of multiple vector-potentials in different sectors of the Hilbert space that correspond to different order of hoppings. We note that this possibility was also identified earlier in the context TMDCs \cite{Rostami2013}. We also discuss building a model for strain corrections from Hilbert space projections from a large dimensional system which becomes relevant when trying to build a model for a low energy Hamiltonian of a system.

We apply our results to various Bravais and non-Bravais lattices (which include square, triangular, graphene, Lieb, Kagom\'e and super-Honeycomb lattices) where we corroborate our results with an explicit calculation. The celebrated 2D material, graphene, in this context, is shown to be a special case which permits the interpretation of the strain correction as vector-potential. Finally, we consider the case of straining a bilayer graphene where we identify multiple vector-potentials arising from in-plane and out-of-plane hoppings. In its low-energy projected subspace, we are able to identify a strain dependent energy scale that separates regions where the presence of multiple vector-potentials is essential and where a single potential is sufficient.

The rest of the text is organized as follows. To model strain, a coarse graining of the system in the presence of slow variations is needed and is a standard practice\cite{sundaram_wave-packet_1999}. However, for completeness, we review the procedure in Sec. \ref{Sec:CoarseGrain} as the formulas from this section will be used in the rest of the text. In this section we will arrive at the general form of strain corrections to the parameters of the lattice. In Sec. \ref{Sec:LowEnergyNeighboringGrains} we use these parametric corrections to find the strain correction to the matrix elements of any 2D Hamiltonian expressed within a tight-binding model. Here, we use group theory to interpret the corrections as scalar- and vector-potentials. In Sec. \ref{Sec:Examples} we demonstrate the validity of our results for prototypical systems involving Bravais and non-Bravais square and triangular lattices, and discuss systems with larger Hilbert space sizes. In Sec. \ref{Sec:3D} we apply the theory to bilayer graphene to demonstrate how these concepts can be applied straightforwardly to systems that are being studied actively. In Sec. \ref{Sec:Conclusions} we summarize the main statements and present a discussion on the prospects of this theory towards search for new physics. Finally, in the Appendix we provide details of explicit calculations of strain corrections for the examples considered in the text.

\section{Review of coarse-graining of strain}\label{Sec:CoarseGrain}
The primary effect of strain on a solid is to displace its atomic sites from location $\br^0_i$ (where $i$ enumerates the sites) to $\br_i$. The standard way (see e.g. Refs. \cite{CastroNeto2009,Naumis2017}) to describe this deformation is to introduce a deformation field ${\boldsymbol{\mathcal D}}(\br_i^0)$ which describes the deformation of the atom that was at $\br^0_i$. The new location can then be written as $\br_i=\br_i^0+{\boldsymbol{\mathcal{D}}}(\br^0_i)$. Although one only needs $\boldsymbol{\mathcal{D}}$ at the discrete lattice sites $\br_i$, the field is well defined everywhere and varies slowly over many lattice constants when an external strain is applied. This allows us to treat $\br_i$ as a continuous variable $\br$, making $\boldsymbol{\mathcal D}$ differentiable. Then, a vector connecting two points at $\br^{a,0}$ and $\br^{b,0}$ in the lattice changes as 
\beq\label{eq:DistanceCorrection}
\underbrace{\br^{a}-\br^{b}}_{\Delta{\bf l}^{ab}}=\underbrace{\br^{a,0}-\br^{b,0}}_{\Delta{\bf l}^{ab,0}} + \boldsymbol{\mathcal D}(\br^{a,0})-\boldsymbol{\mathcal D}(\br^{b,0}).
\eeq
For small enough distances (still over a few lattice constants), this can be written in component form as
\bea\label{eq:DistanceCorrection2}
\Delta{l}^{ab}_\alpha&=&\Delta{l}^{ab,0}_\alpha + \underbrace{\partial_\beta\mathcal D_\alpha}_{u_{\alpha\beta}}\Delta l^{ab,0}_\beta+...\nn\\
&=&(\delta_{\alpha\beta}+u_{\alpha\beta})\Delta l_\beta^{ab,0}+...,
\eea
where $\alpha,\beta\in\{x,y\}$ and $u_{\alpha\beta}$ is the displacement-gradient tensor as introduced in the elasticity theory, which can also be expressed in terms of the symmetric linear strain tensor and the anti-symmetric rotational strain tensor \cite{de_juan_gauge_2013,ramezani_masir_pseudo_2013}. The `$+...$' refers to second and higher order corrections which we don't account for in this work. Throughout the text we follow the Einstein convention where repeated pair of indices are summed. If the points $a$ and $b$ are in the vicinity of each other with an average location $\br$, then the above equation can be written in a position dependent fashion as (ignoring higher order corrections in the deformations)
\bea\label{eq:DistanceCorrection3}
\Delta{l}^{ab}_\alpha(\br)&=&[\delta_{\alpha\beta}+u_{\alpha\beta}(\br)]\Delta l_\beta^{ab,0}(\br).
\eea
It is important to understand that the above equation presents a strain induced correction to the separation between any two points $a,b$ in the vicinity of $\br$. These points could be the adjacent lattice sites, or nearest neighbor (nn) sites, or next nn site, etc. The $\delta_{\alpha\beta}+u_{\alpha\beta}(\br)$ factor provides the scaling tensor for the length correction due to strain.

An applied strain can be of three types: (i) Uniform - In this case, which is the most common scenario, $u_{\alpha\beta}(\br)$ is a constant. While this reduces the rotational symmetry of the lattice, it  preserves the translational ones. In this case, if Eq. (\ref{eq:DistanceCorrection3}) is applied to the separation between lattice sites, then it is evident that the lattice constant is simply scaled by strain. (ii) Periodic - This type breaks the translational symmetry locally, but restores it at a larger scale leading to formation of superstructures. This type of strain is more natural in systems on a substrate or in twisted 2D structures \cite{Kogl2023, Banerjee2020}. (iii) Non-uniform - In this case (realized e.g. in Ref. \cite{Levy2010}), the translational symmetry is broken at all scales. The type(i) strain preserves the number of periodic units in the sample but alters the lattice translation vectors. This means that the BZ has the same number of quantum states but is merely stretched/squeezed due to changes in the translation vectors. Type(ii) deformations lead to enlarging of the unit cell and leads to zone folding effects. Type(iii) strain can be intractable in general.

However, imagine a slowly varying $u_{\alpha\beta}(\br)$ such that the system could be broken down into piece-wise type(i) system, which we shall call `grains'. If $\eta_{\rm tol}$ were to denote the tolerance below which we ignore changes in the tensor $u_{\alpha\beta}$, then a possible grain size around the grain coordinate $\bR_n$, defined as $l_g\equiv|\bR_n-\br_{\rm bound}|$, could be obtained from 
\bea\label{eq:grainsize}
\frac{|\partial_\br u_{\alpha\beta}(\bR_n)|}{|u_{\alpha\beta}(\bR_n)|}l_g<\eta_{\rm tol}.
\eea
Here $\br_{\rm bound}$ would be some locus of points around $\bR_n$ that would mark the boundary of the grain. $\eta_{\rm tol}$ is usually set such that the resulting grain is large enough to validate the use of periodic boundary conditions leading to formation of a reasonably continuous Bloch band structure within each grain. For such slow variations, one can impose $u_{\alpha\beta}(\br)\rightarrow u_{\alpha\beta}(\bR_n),~\forall\br\in$ neighbourhood of $\bR_n$. We refer to this procedure as coarse-graining of the strain. For a large enough system the coarse-grained $\bR_n$ can also be treated continuously as $\bR$. Note that if the gradient in strain is very small or absent, the grain size $l_g$ can be as large as the sample.

\begin{figure}[ht]
    \centering
    \includegraphics[width=0.85\textwidth]{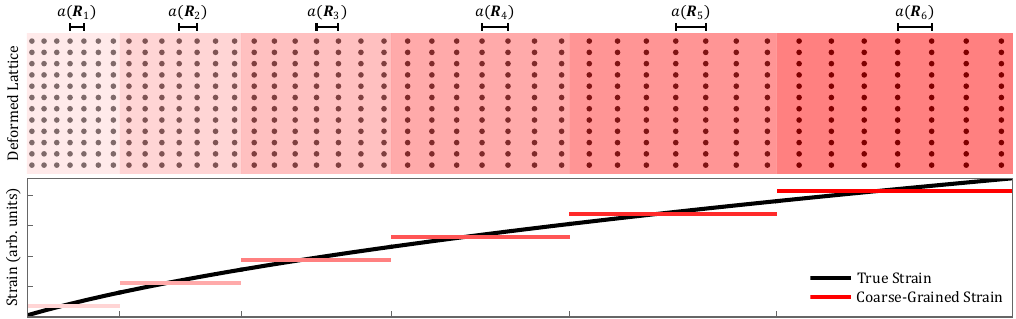}
    \caption{\textbf{Coarse-graining of non uniform strain}. The shades track the strain value across the sample that has been uniaxially stretched by deformation potential $\boldsymbol{\mathcal D}(\br)\propto x^2\hat x$. The variations within a grain that are below a set tolerance (exaggerated here for visual clarity) can be ignored and assigned a single uniform strain value $\e_{\alpha\beta}(\bR_i)$. A wavepacket spanning a few grains will experience a slowly varying strain field. A grain at $\bR_i$ can be seen as having a constant lattice constant $a(\bR_i)$. For large samples, $\bR_i$ is promoted to a continuous variable $\bR$.} \label{Fig:2}
\end{figure}

Our system can now be seen as a juxtaposition of subsystems (grains) that are parameterized by $\bR$ (see Fig. \ref{Fig:2}). The grain Hamiltonian can be written in the following manner (we will ignore spin throughout this work as mechanical strain without spin-orbit coupling will not couple to the spin degree of freedom):
\beq\label{eq:Hamiltonian}
\hat H(\bR,\bk)=\sum_{\bk}^{(\bR)} h_{ab}(\bR,\bk)\hat c^\dag_{a,\bk} \hat c_{b,\bk},
\eeq
where $a,b$ span the Hilbert space of the Hamiltonian. The sum is over the number of $\bk$ vectors in the first BZ, which is just the number of unit cells in the grain. This sum depends implicitly on $\bR$ through the modifications of the lattice vectors (and thus the BZ size) in each grain. The more explicit dependence on $\bR$ appears in the matrix elements $h_{ab}$ that change due to the altered distances between respective atomic centres due to $u_{\alpha\beta}(\bR)$. These modifications lead to grain dependent band structures $E(\bR,\bk)$ and wavefunctions. If $\hat H$ is modelled within a tight-binding model, we can express the matrix element as a sum over hoppings to neighbors of different orders:
\beq\label{eq:Hopping}
h_{ab}(\bR,\bk)=\sum_{n}^{\rm orders}\left[\sum_{j_n}^{\rm neighbors} t^{(j_n)}_{a_nb_n}(\bR) e^{-i\bk\cdot\boldsymbol{\delta}_{a_nb_n}^{(j_n)}(\bR)}\right].
\eeq
Here, $t^{(j_n)}_{a_nb_n}$ is the real space hopping element between $a,b$ elements of the Hilbert space, but from the $n^{\rm th}$ order of the bonds. An order is identified by a particular distance between the charge centers as well as the type of charge centers. When there are different types of charge centres in the system at the same distance, there is a freedom of choice between including them in the same order or keeping them as separate. The index $j_n$, on the other hand, runs over the co-ordination number of the neighbors of a given order. By definition,  $|\boldsymbol \delta^{(j_n)}_{a_nb_n}|$ is the same for a given $n$ which we may denote as $\delta_{a_nb_n}$ (e.g. see grey and green bonds in Fig. \ref{Fig:1}a). The hopping $t^{(j_n)}_{a_nb_n}$ is usually independent of $j_n$, unless the hoppings are made to be phase sensitive. Both $t^{(j_n)}_{a_nb_n}$ and $\boldsymbol{\delta}^{(j_n)}_{a_nb_n}$ depend on $\bR$ via the displacement-gradient tensor $u_{\alpha\beta}(\bR)$. This dependence for $\boldsymbol{\delta}_{a_nb_n}^{(j_n)}(\bR)$ can be written following Eq. (\ref{eq:DistanceCorrection3}) as:
\beq\label{eq:LengthConnection}
\delta^{(j_n)}_{a_nb_n,\alpha}(\bR)=[\delta_{\alpha\beta}+u_{\alpha\beta}(\bR)]\delta^{(j_n)}_{a_nb_n,\beta},
\eeq
where $\delta^{(j_n)}_{a_nb_n,\beta}$ is the unstrained vector. Henceforth, unless otherwise mentioned, we will work with $n=1$ and also drop this subscript for brevity and with the understanding that the arguments can be readily extended to any order $n$. The deformed distance, on the other hand, is written as
\bea\label{eq:MagCorrections}
|{\boldsymbol \delta}^{(j)}_{ab}(\bR)|^2&=&\delta^2_{ab}\left(1+\underbrace{[u_{\alpha\beta}(\bR)+u_{\beta\alpha}(\bR)+u_{\gamma\alpha}(\bR)u_{\gamma\beta}(\bR)]}_{2\e_{\alpha\beta}(\bR)}\frac{\delta^{(j)}_{ab,\alpha}\delta^{(j)}_{ab,\beta}}{\delta^2_{ab}}\right)\nn\\
\Rightarrow |{\boldsymbol \delta}^{(j)}_{ab}(\bR)|&=&\delta_{ab}\left(1+\e_{\alpha\beta}(\bR)\frac{\delta^{(j)}_{ab,\alpha}\delta^{(j)}_{ab,\beta}}{\delta^2_{ab}}\right)+\mathcal{O}(u^2_{\alpha\beta}),
\eea
where $\epsilon_{\alpha\beta}(\bR)$ is the strain tensor. In the last line in Eq. (\ref{eq:MagCorrections}), we have restricted ourselves to linear effects of deformation (see e.g. Refs. \cite{ramezani_masir_pseudo_2013,Oliva-Leyva_2017} for non-linear effects which will also include effects of out-of-plane deformations). The hopping element depends on the distance between the charge centres of the states $a,b$ as $t^{(j)}_{ab}(\bR)=t^{(j)}_{ab}e^{-\zeta(\delta_{ab}(\bR)/\delta_{ab}-1)}$, where $\zeta$ is the electron Gr\"uneisen parameter \cite{Mohiuddin2009} for the bond. The parameter $\zeta$ is controlled by the nature of the orbitals involved in bond formation and hence will quite generally depend on the order of the bonds considered. We can then calculate the modification in the hopping element as:
\bea\label{eq:HopCorrections}
t^{(j)}_{ab}(\bR)&\approx&t^{(j)}_{ab}\left(1-\zeta\e_{\alpha\beta}(\bR)\frac{\delta^{(j)}_{ab,\alpha}\delta^{(j)}_{ab,\beta}}{\delta^2_{ab}}\right)\nn\\
&=&t^{(j)}_{ab}[1-\zeta\e_{\alpha\beta}(\bR)d^{(j)}_{ab,\alpha}d^{(j)}_{ab,\beta}],
\eea
where we have introduced the unit vectors $d^{(j)}_{ab,\alpha}\equiv \delta_{ab,\alpha}^{(j)}/\delta_{ab}$ for brevity. We make an observation here that distances are modified by the symmetric strain tensor $\e_{\alpha\beta}$, but vectors are modified by the displacement-gradient tensor $u_{\alpha\beta}$ which is also sensitive to local rotations.

\section{Strain corrections to the low energy Hamiltonian in a grain}\label{Sec:LowEnergyNeighboringGrains}
Having calculated the strain corrections to the lattice parameters (the nn vectors and the hopping elements), we can now get the leading correction to the Hamiltonian at any point $\bK$ in the BZ as:
\bea\label{eq:strainExpansion}
h_{ab}(\bR,{\bK})&\approx& \sum_j t^{(j)}_{ab}e^{-i{\bK}\cdot\boldsymbol{\delta}_{ab}^{(j)}}[1-iK_\alpha u_{\alpha\beta}(\bR)\delta^{(j)}_{ab,\beta}-\zeta\e_{\alpha\beta}(\bR)d^{(j)}_{ab,\alpha}d^{(j)}_{ab,\beta}]\nn\\
&=&\underbrace{\sum_j t^{(j)}_{ab}e^{-i{\bK}\cdot\boldsymbol{\delta}_{ab}^{(j)}}}_{h^0_{ab}(\bK)}+\underbrace{K_\alpha u_{\alpha\beta}(\bR)\sum_j t^{(j)}_{ab}e^{-i{\bK}\cdot\boldsymbol{\delta}_{ab}^{(j)}}[-i\delta^{(j)}_{ab,\beta}]}_{s^D_{ab}(\bR,\bK)}+\underbrace{\zeta\e_{\alpha\beta}(\bR)\sum_j t^{(j)}_{ab}e^{-i{\bK}\cdot\boldsymbol{\delta}_{ab}^{(j)}}[-d^{(j)}_{ab,\alpha}d^{(j)}_{ab,\beta}]}_{s^H_{ab,\parallel}(\bR,\bK)}.
\eea
Here, by $h^0_{ab}(\bK)$ we denote the unstrained Hamiltonian. The strain correction terms to leading order are: $s^D_{ab}(\bR,\bK)$ from the displacements ($D$) of the lattice, and $s^H_{ab,\parallel}(\bR,\bK)$ from the change in hoppings ($H$). The former involves the displacement-gradient tensor, while the latter, the strain tensor. Consider also the expansion of the unstrained $h^0_{ab}$ in a small momentum $\bq$ about the same point $\bf K$ in the BZ to second order:
\bea\label{eq:qExpansion}
h^0_{ab}(\bR,{\bK}+\bq)&=&h^0_{ab}({\bK}) + \underbrace{\sum_j t^{(j)}_{ab}(\bR)e^{-i{\bK}\cdot\boldsymbol{\delta}_{ab}^{(j)}(\bR)}[-i\delta_{ab,\alpha}^{(j)}(\bR)]}_{v_{ab,\alpha}(\bR,{\bK})}q_\alpha+\underbrace{\sum_j t^{(j)}_{ab}(\bR)e^{-i{\bK}\cdot\boldsymbol{\delta}_{ab}^{(j)}(\bR)}[-\delta_{ab,\alpha}^{(j)}(\bR)\delta_{ab,\beta}^{(j)}(\bR)]}_{I_{ab,\alpha\beta}(\bR,{\bK})}\frac{q_\alpha q_\beta}2,\nn\\
\eea
where $v_{ab,\alpha}(\bR,{\bK})\equiv \partial_{k_\alpha}h^0_{ab}(\bR,\bK)$ is the $\alpha^{\rm th}$ component of velocity matrix element and $I_{ab,\alpha\beta}\equiv\partial_{k_\alpha}\partial_{k_\beta}h^0_{ab}(\bR,\bK)$ is the $\alpha,\beta^{\rm th}$ component of the inverse mass tensor matrix element between states $a,b$ at point $\bf K$ in the BZ in the grain at $\bR$. Upon inspecting Eqs. (\ref{eq:strainExpansion}) and (\ref{eq:qExpansion}) we see that
\bea\label{eq:StrainClass}
s^D_{ab}(\bR,\bK)&=&K_\alpha u_{\alpha\beta}(\bR)v_{ab,\beta}(\bR,{\bK}),\nn\\
s^H_{ab,\parallel}(\bR,\bK)&=&\zeta\e_{\alpha\beta}(\bR)I_{ab,\alpha\beta}(\bR,\bK)/\delta_{ab}^2,
\eea
That is, the displacement induced strain correction couples to the Hamiltonian via the velocity and the hopping induced strain correction couples via the inverse mass tensor. Thus, the strained low energy Hamiltonian can be written as
\bea\label{eq:FullExpansion}
h_{ab}(\bR,{\bK}+\bq)&=&h^0_{ab}({\bK}) + \left[q_\beta+K_\alpha u_{\alpha\beta}(\bR)\right]v_{ab,\beta}({\bK})+\zeta\e_{\alpha\beta}(\bR)I_{ab,\alpha\beta}(\bK)/\delta_{ab}^2+\mathcal{O}(q^2,u_{\alpha\beta}^2, qu_{\alpha\beta}).
\eea
The velocity and inverse mass are strain independent when considering corrections to the order stated above. This is valid for small strain and low carrier densities (leading to small Fermi momentum and hence small $q$). Velocity will acquire strain corrections from $\mathcal{O}(qu_{\alpha\beta})$ terms \cite{ramezani_masir_pseudo_2013,oliva-leyva_generalizing_2015} and the inverse mass will acquire strain corrections from $\mathcal{O}(u^2_{\alpha\beta})$ terms which are not considered in this work. 

Because the velocity and the inverse mass tensor only have in-plane components, the strain corrections in Eq. (\ref{eq:StrainClass}) will only pick up contributions from in-plane displacements. But the bonds are not restricted to be in-plane even if the lattice is 2D (e.g. the unit cell may have a 3D structure like in the TMDCs or multilayer systems). If a given bond were to have an out-of-plane component, then the hopping correction from Eq. (\ref{eq:strainExpansion}) would split into three contributions (see also chart in Fig. \ref{Fig:1}c): one from the in-plane projections of the bonds which couples to the inverse mass tensor; one from when one of the $\boldsymbol{\delta}^{(j)}$ contributes via the out-of-plane ($z$) component and the other via the in-plane component; and the third from when both $\boldsymbol{\delta}^{(j)}$'s contribute via its $z$ component. Explicitly, we can write this correction as
\bea\label{eq:zstrain}
s^{H}_{ab,\perp}(\bR,\bK)&\equiv&\sum_j t^{(j)}_{ab}e^{-i{\bK}\cdot\boldsymbol{\delta}_{ab}^{(j)}}\zeta\left\{\e_{\alpha z}(\bR)[-d^{(j)}_{ab,\alpha}d^{(j)}_{ab,z}]+\e_{z\beta}(\bR)[-d^{(j)}_{ab,z}d^{(j)}_{ab,\beta}]\right\}+
\sum_j t^{(j)}_{ab}e^{-i{\bK}\cdot\boldsymbol{\delta}_{ab}^{(j)}}\zeta\e_{zz}(\bR)[-d^{(j)}_{ab,z}d^{(j)}_{ab,z}]\nn\\
&=&-2i\zeta\e_{\alpha z}(\bR)v_{ab,\alpha}(\bK)\delta_{ab,z}/\delta_{ab}^2-\zeta\e_{zz}(\bR){h^0_{ab}}(\bK)\delta^{2}_{ab,z}/\delta_{ab}^2.
\eea
Unlike $\hat s^H_\parallel$, $\hat s^{H}_\perp$ is not expressible in terms of the inverse mass, but the velocity and the scalar matrix $h^0_{ab}$. The linear-in-$\delta_{ab,z}$ correction comes in the form $i\delta_{ab,z}$. Since $\delta_{ab,z}=-\delta_{ba,z}$, the $i$ must be present to ensure the Hermitian property of the Hamiltonian. Combining all the in- and out-of-plane deformations, we can write
\bea\label{eq:FullExpansion2}
h_{ab}(\bR,{\bK}+\bq)&=&h^0_{ab}({\bK})+s^D_{ab}(\bR,\bK)+s^H_{ab,\parallel}(\bR,\bK)+s^H_{ab,\perp}(\bR,\bK)\nn\\
~&=&\left[1-\zeta\e_{zz}(\bR)\delta^2_{ab,z}/\delta_{ab}^2\right]h^0_{ab}({\bK})+\left[q_\beta+K_\alpha u_{\alpha\beta}(\bR)-2i\zeta\e_{\beta z}(\bR)\delta_{ab,z}/\delta_{ab}^2\right]v_{ab,\beta}({\bK})+\nn\\
&&~\zeta\e_{\alpha\beta}(\bR)I_{ab,\alpha\beta}(\bK)/\delta_{ab}^2+\mathcal{O}(q^2,u_{\alpha\beta}^2, qu_{\alpha\beta}).
\eea
Equation (\ref{eq:FullExpansion2}) represents \textit{our first general result} which expresses the leading order strain correction to Hamiltonian of any 2D system in terms of the already known band structure parameters such as the velocity and the inverse mass matrices.

\subsection{Strain induced potentials}
If one constructs a wavepacket out of the Fermi level states of this low energy Hamiltonian, the packet would span a few grains and would evolve according to a Schr\"odinger equation obtained from Eq. (\ref{eq:FullExpansion2}) by replacing $q_\alpha\rightarrow -i\partial_{R_\alpha}$. This is often referred to as the continuum limit. This continuum limit Hamiltonian then feels a perturbation from strain as various potentials. The perturbation that is $\propto\hat h^0$ serves as a scalar-potential, while the ones that are $\propto\hat v_{\alpha}$ serve as a vector-potential. From Eqs. (\ref{eq:StrainClass}), (\ref{eq:zstrain}) and (\ref{eq:FullExpansion2}), it is evident that the $\hat s^D$ and $\hat s^H_\perp$ parts of the strain corrections already have terms that serve as strain induced scalar/vector-potentials. However, the interpretation of the potential arising from $\hat s^H_\parallel$ is not clear, and finding it is the purpose of this section.

First, note that the potential-like interpretation is only possible if all components of the wavepacket feel the same potential. If the Fermi surface is large, then the wavepacket that is formed out of states at the Fermi level would comprise of many $\bK$ components and hence many different potentials. This would lead to a convolution of the potential with the wavepacket. However, if we were to consider a lightly doped system at some singular point, like the extremum of a band or a point of degeneracy, then the wavepacket would only involve states around one $\bK$ point (in the form $\bK+\bq$, with small $\bq$), allowing for a single effective potential for the wavepacket. There may be multiple singular points at the Fermi level, in which case the wavepacket may be seen as a superposition of the packets from countably finite number of singular points, each with its own potential. Once, we have identified our singular point of interest, $\bK$, we can decompose the strain correction $\hat s^D+\hat s^H_{\parallel}+\hat s^H_{\perp}$ in terms of the generators of the Hilbert space. Now, a $d_H$ dimensional Hilbert space has $d_H^2-1$ generators plus the identity. However, group theoretical considerations guarantee that we only need a handful of them for the decomposition.

To see this, let us remind ourselves that any Hamiltonian $\hat h^0$, although a matrix in the Hilbert space, is a scalar (which is invariant under all actions of the point group of the lattice). Accordingly, the velocity matrix must transform as a vector, and the inverse mass matrix as a rank 2 tensor. The only way a velocity may appear in the Hamiltonian is if it is contracted with another vector (which is usually the momentum $q_\alpha$). Likewise, the inverse mass must contract with another tensor. However, if we consider the Hamiltonian matrix element at point $\bK$ from Eq. (\ref{eq:Hopping}) that has the form $\sum_jt^{(j)}e^{-i\bK\cdot\boldsymbol{\delta}_{ab}^{(j)}}$, it would be invariant only under the operations belonging to the wavevector group at $\bK$ (subgroup of the point group that leaves the $\bK$ point invariant). In fact, and this is often not emphasized, the symmetry group of $h_{ab}(\bK)$ that is formed out of a specific order of hoppings, is the common subgroup of the wavevector group at $\bK$ and the group of symmetries that leave the $ab$ subspace bonds invariant. We refer to this as the \textit{bond-wavevector group}. 

The matrix structure of $h^0_{ab}$ that is a scalar, must then belong to the fully symmetric irrep of the bond-wavevector group. For such a term to exist, we must have a scalar generator of the Hilbert space. This is guaranteed if $h^0_{ab}\neq0$. The velocity matrix element from Eq. (\ref{eq:qExpansion}) has a linear form $\sum_jt^{(j)}e^{-i\bK\cdot\boldsymbol{\delta}^{(j)}_{ab}}\delta^{(j)}_{ab,\alpha}$ characterized by the single index $\alpha$. This must transform as a vector in the bond-wavevector group and thus belongs to the irrep that contains the functions $x,y$ in its character table. For such a term to exist there must be a pair of generators of Hilbert space serving as 2D vectors that can couple to these vector forms. If we cannot find such matrices, then there can be no velocity terms in the Hamiltonian. Finally, the inverse mass tensor is a quadratic form $\sum_jt^{(j)}e^{-i\bK\cdot\boldsymbol{\delta}_{ab}^{(j)}}\delta^{(j)}_{ab,\alpha}\delta^{(j)}_{ab,\beta}$ which must transform as a tensor of rank 2, i.e. the $x^2,y^2,xy$ functions of the character table. These quadratic functions can further be decomposed into different irreps. In Table \ref{Groups_table}, we can compare the distribution of various linear and quadratic functions in the character tables of different groups that will be described in the text. Since $\hat s^H_\parallel$ is a quadratic form, it can create a potential in every irrep that contains the quadratic functions. However, the presence of a potential in a particular irrep also relies on the availability of generators of the Hilbert space that can represent that irrep. If no such generators can be found, then potentials of that irrep will not exist. These arguments apply not only to strain but also to $\bq$-expansions of the unstrained Hamiltonians at any given $\bK$-point.

Returning back to our problem, we now see that since the strain corrections involve scalars, linear and quadratic forms, we only need those generators that correspond to these (which are at most 5 in number in 2D: $ x,y,x^2,y^2,xy$) and not necessarily to the full Hilbert space. If the linear and quadratic forms were to share an irrep, then the same generator can incorporate both types of corrections. This only happens in the bond-wavevector groups $C_1$ (the trivial group), $C_{1v}$ (which has identity and mirror as the only operations) and $C_{3v}$, where the quadratic functions fall under the vector irreps (see Table \ref{Groups_table})\footnote{If one considers the presence of a horizontal mirror, commonly denoted as $\sigma_h$, one would have to work with $D_{nh}$ groups. Since $C_{nv}$ groups are subgroups of $D_{nh}$ groups, the same ideas will apply.}. 

\begin{table}
    \centering
    \begin{tabularx}{0.5\textwidth} {|>{\centering\arraybackslash\hsize=.5\hsize}X|>{\centering\arraybackslash\hsize=.5\hsize}X|>{\centering\arraybackslash}X|>{\centering\arraybackslash}X|}
         \hline
         \multicolumn{2}{|c|}{Group/IRREPs} & Linear & Quadratic \\\hhline{|====|}
         $C_1$ & $A_1$ & $x,y$ & $x^2+y^2,x^2-y^2,xy$ \\\hhline{|====|}
         \multirow{2}{*}{$C_{1v}$} & $A_1$ & $x$ & $x^2+y^2,x^2-y^2$ \\\cline{2-4}
                                   & $A_2$ & $y$ & $xy$\\\hhline{|====|}
         \multirow{4}{*}{$C_{2v}$} & $A_1$ & $-$ & $x^2+y^2,x^2-y^2$ \\\cline{2-4}
                                   & $A_2$ & $-$ & $xy$\\\cline{2-4}
                                   & $B_1$ & $x$ & $-$\\\cline{2-4}
                                   & $B_2$ & $y$ & $-$\\\hhline{|====|}
         \multirow{3}{*}{$C_{3v}$} & $A_1$ & $-$ & $x^2+y^2$ \\\cline{2-4}
                                   & $A_2$ & $-$ & $-$\\\cline{2-4}
                                   & $E$ & $(x,y)$ & $(x^2-y^2,xy)$ \\\hhline{|====|}
         \multirow{5}{*}{$C_{4v}$} & $A_1$ & $-$ & $x^2+y^2$ \\\cline{2-4}
                                   & $A_2$ & $-$ & $-$\\\cline{2-4}
                                   & $B_1$ & $-$ & $x^2-y^2$ \\\cline{2-4}
                                   & $B_2$ & $-$ & $xy$\\\cline{2-4}
                                   & $E$ & $(x,y)$ & $-$ \\\hhline{|====|}
         \multirow{6}{*}{$C_{6v}$} & $A_1$ & $-$ & $x^2+y^2$ \\\cline{2-4}
                                   & $A_2$ & $-$ & $-$\\\cline{2-4}
                                   & $B_1$ & $-$ & $x^2-y^2$ \\\cline{2-4}
                                   & $B_2$ & $-$ & $xy$\\\cline{2-4}
                                   & $E_1$ & $(x,y)$ & $-$ \\\cline{2-4}
                                   & $E_2$ & $-$ & $(x^2-y^2,xy)$ \\\hline
        \end{tabularx}
    \caption{Character table functions of the selected symmetry groups}
    \label{Groups_table}
\end{table}

There isn't any universal scheme to classify generators into different irreps. Thus, the same matrices can serve as generators of different irreps in different groups. But once a bond-wavevector group is fixed, the generators' association with a particular irrep is also fixed, and they can be deduced by looking at the matrix structure of the unstrained Hamiltonian and its $\bq$ expansions. In a Hilbert space of dimension $d_H$, these generators, say $\hat M^r$, have the property that Tr$[\hat M^r\hat M^{r'}]=d_H\delta_{rr'}$. This allows us to uniquely project the corrections onto the generators. In general, there will be a generator $\hat M^0\propto\hat h^0$ projecting on to which will yield a scalar-potential. There may also exist generators $\hat M^{\alpha}\propto\hat v_{\alpha}$ projecting on to which will yield a vector-potential. Similarly, one can identify potentials in any other irrep $r$ of the bond-wavevector group. To this effect, observe that $\hat s^D$ only has projection on to the velocity indicating that it can \textit{only} give rise to a vector-potential as prescribed in the minimal coupling approach:
\bea\label{eq:ADDef}
e\mathcal A^D_\beta(\bR,{\bK})&\equiv& {\rm Tr}[\hat s^D(\bR,\bK)\hat v_\beta(\bK)]/v^2_\beta,~\text{where }v_\beta^2\equiv {\rm Tr}[\hat v_{\beta}(\bK)\hat v_{\beta}(\bK)],\nn\\
&=&K_\alpha u_{\alpha\beta}(\bR).
\eea
Here $e$ is the fundamental charge and has been introduced in the definition to make $\mathcal A^{D}_\alpha$ have the same dimensions as the electromagnetic vector-potential. 

Next, observe that $\hat s^H_\perp$ also has a component that is $\propto -i\delta_{ab,z}v_{ab,\alpha}$. Since $\delta_{ab,z}$ switches sign between the orderings $ab$ and $ba$, it will be beneficial to write it as $\delta_{ab,z}=\ve_{ab}|\delta_{ab,z}|$, where $|\delta_{ab,z}|$ is nothing but the inter-plane distance between the bonds and $\ve_{ab}$ is an anti-symmetric matrix with unit entries (not to be confused by the strain tensor $\e_{\alpha\beta}$). We then introduce another pair of generators such that $\bar v_{ab,\alpha}\equiv -i\varepsilon_{ab}v_{ab,\alpha}$. Note that Tr$[\hat v_{\alpha}\hat{\bar v}_\alpha]=0$ ensuring that $\hat{\bar v}_\alpha,\hat v_\alpha$ are different generators. However, the set of $\bar v_{ab,\alpha}$ can still map back to the set of the velocity generators $v_{ab,\alpha}$ with rearranged indices $\alpha$ (e.g. $-i\varepsilon_{ab}\sigma_{ab,x}=\sigma_{ab,y}$ and $-i\varepsilon_{ab}\sigma_{ab,y}=-\sigma_{ab,x}$). This case can arise when a third pseudo-vector (e.g. $\hat z$) is present in the problem that permits the scalar $\hat{\mathbf{v}}\times\bq\cdot\hat z$ to also be present. In such a case, we can extract vector-potentials simply by using the velocity generators:
\bea\label{eq:ADDef2}
e\mathcal A^{H,\perp}_\beta(\bR,{\bK})&\equiv&{\rm Tr}[s^H_\perp(\bR,\bK)\hat{v}_\beta(\bK)]/v_\beta^2.
\eea
If the new generators don't map back to themselves, then they would represent some other irrep $r$ of the system. In such cases, we can normalize the new generator $\hat{\bar v}_{\alpha}$ to $\hat M^\alpha$ such that ${\rm Tr}[\hat M^\alpha.\hat M^\alpha]=d_H$. We can lump these irreps together with the scalar form and label them by an index $r$. These generators can then be used to extract potentials in the system as:
\bea\label{eq:phiDef}
-e\varphi^{H,\perp}_r(\bR,{\bK})&\equiv&{\rm Tr}[\hat s^H_\perp(\bR,\bK)\hat M^r(\bK)]/d_H.
\eea
For $r=0$ we have the scalar-potential of the form
\bea\label{eq:phiDef2}
-e\varphi^{H,\perp}_0(\bR,{\bK})&\equiv&{\rm Tr}[\hat s^H_\perp(\bR,\bK)\hat M^0(\bK)]/d_H\nn\\
&=&-\zeta\e_{zz}(\bR){\rm Tr}[\hat h^0(\bK)\hat M^0(\bK)]\delta^2_{ab,z}/d_H\delta_{ab}^2.
\eea
For the other irreps with $r=x,y$ we get
\bea\label{eq:phiDef3}
-e\varphi^{H,\perp}_x(\bR,{\bK})&\equiv&{\rm Tr}[\hat s^H_\perp(\bR,\bK)\hat M^x(\bK)]/d_H\nn\\
&=&2\zeta\e_{x z}(\bR)\frac{|v_{x}||\delta_{ab,z}|}{\delta_{ab}^2},\nn\\
-e\varphi^{H,\perp}_y(\bR,{\bK})&\equiv&{\rm Tr}[\hat s^H_\perp(\bR,\bK)\hat M^y(\bK)]/d_H\nn\\
&=&2\zeta\e_{yz}(\bR)\frac{|v_{y}||\delta_{ab,z}|}{\delta_{ab}^2},
\eea
where $|v_\alpha|^2={\rm Tr}[\hat v_\alpha^2]/d_H$. 

The potentials from $\hat s^H_\parallel$ corrections are more straightforward and can be deduced from:
\bea\label{eq:Hpotss}
-e\varphi^{H}_r(\bR,\bK)&=&{\rm Tr}[\hat s^H_{\parallel}(\bR,\bK)\hat M^r(\bK)]/d_H,\nn\\
e\mathcal A^{H,\parallel}_\alpha(\bR,\bK)&=&{\rm Tr}[\hat s^H_{\parallel}(\bR,\bK)\hat v_{\alpha}(\bK)]/v_\alpha^2,
\eea
where, as discussed before, the velocity based decomposition only exists for $C_{1v}$ or $C_{3v}$ bond-wavevector groups. Note that when the generators $\{\hat{\bar v}_\alpha\}$ map back to $\{\hat v_\alpha\}$, then $\mathcal A^{H,\perp}$ can be incorporated in the formula for $\mathcal A^{H,\parallel}$ in Eq. (\ref{eq:Hpotss}) by replacing $\hat s^H_\parallel$ with $\hat s^H_\perp$. 

This bond-wavevector group specific decomposition of the strain correction into the scalar/vector-potentials coupled to respective generators is the \textit{second general result of the work}. The specific forms of the generators will be discussed in Sec. \ref{Sec:Examples} on a case-by-case basis. The potentials introduced in this manner are necessarily real owing to the Hermiticity of the generators. In summary, starting from our first result, we may list the following prescription to get the potentials:
\begin{subequations}\label{Potentials}
\bea
h_{ab}(\bR,{\bK}+\bq)&=&h^0_{ab}({\bK})+v_{ab,\alpha}({\bK})q_\alpha+s^D_{ab}(\bR,{\bK})+s^H_{ab,\parallel}(\bR,{\bK})+s^H_{ab,\perp}(\bR,{\bK}),\label{AForm}\\
e\mathcal A^D_{\alpha}(\bR,\bK)&=&{\rm Tr}[\hat s^D(\bR,\bK)\hat v_\alpha(\bK)]/v^2_\alpha=K_\beta u_{\beta\alpha}(\bR),\label{AFormAD}\\
-e\varphi^{H,\perp}_r(\bR,{\bK})&=&{\rm Tr}[\hat s^H_\perp(\bR,\bK)\hat M^r(\bK)]/d_H,\label{AFormPhiz}\\
e\mathcal A^{H,\perp}_\alpha(\bR,{\bK})&=&{\rm Tr}[s^H_\perp(\bR,\bK)\hat{v}_\alpha(\bK)]/v_\alpha^2,\label{AFormAHz}\\
-e\varphi^{H}_r(\bR,\bK)&=&{\rm Tr}[\hat s^H_{\parallel}(\bR,\bK)\hat M^r(\bK)]/d_H,\label{AFormPhir}\\
e\mathcal A^{H,\parallel}_\alpha(\bR,\bK)&\stackrel{C_{1v}/C_{3v}}{=}&{\rm Tr}[\hat s^H_{\parallel}(\bR,\bK)\hat v_{\alpha}(\bK)]/v_\alpha^2.\label{AFormAH}
\eea
\end{subequations}
The forms of $\hat s^D,\hat s^H_\parallel,\hat s^H_\perp$ are given in Eqs. (\ref{eq:StrainClass}) and (\ref{eq:zstrain}). For a chosen order of bonds, the velocity is obtained from $\hat v_\alpha=\partial_{k_\alpha}\hat H$ and $\hat{\bar v}_\alpha$ is obtainable from $\hat v_\alpha$ [see discussion before Eq. (\ref{eq:ADDef2})]. $\hat M^r$'s are generators in the non-vector irreps of the bond-wavevector group that contain the quadratic functions, with $r=0$ being the scalar irrep. As Eq. (\ref{AFormAH}) indicates, $\boldsymbol{\mathcal A}^{H,\parallel}$ only exists for bond-wavevector group being $C_{1v}$ or $C_{3v}$. Observe that $\boldsymbol{\mathcal A}^D$ is only dependent on the lattice and not on the details of the Hilbert space. The set of Eqs. (\ref{Potentials}) is the mathematical version of the chart in Fig. \ref{Fig:1}c.

\subsection{Universal considerations of strain induced potentials}\label{Subsec:UCon}
\paragraph*{Time reversal symmetry:} Since we are dealing with mechanical deformations of the lattice, time reversal symmetry (TRS) must be preserved if already present in $h^0_{ab}$. This imposes the condition $h_{ab}(\bR,{\bK}+\bq)=h^*_{ab}(\bR,{-\bK}-\bq)$. Since $\bK$ is a singular point, we would need a pair of singular points $(\bK,-\bK)$ to preserve TRS, unless these points are related by translations of the reciprocal space. Since $v_{ab,\alpha}(-{\bK})=-v^*_{ab,\alpha}(\bf K)$ from Eq. (\ref{eq:qExpansion}), and $q_\alpha\rightarrow-q_\alpha$ under TRS, the Hamiltonian pair at $(\bK,-\bK)$ is manifestly invariant under TRS in our expressions. Translating the TRS requirements to the potentials (which are real by construction) we get
\bea\label{eq:pseudocond}
\mathcal A_\alpha(\bR,-\bK)&=&-\mathcal A_\alpha(\bR,\bK),\nn\\
\varphi_r(\bR,-\bK)&=&\varphi_r(\bR,\bK),
\eea
where $\mathcal A_\alpha$ and $\varphi_r$ represent the total vector-potentials and potentials in other irreps from all sources, respectively.

\paragraph*{Pseudo electric and magnetic fields from strain induced potentials:} If the potentials were to vary smoothly from grain to grain, then from the vector-potential identified above, one can define a pseudo magnetic field with only one component directed out-of-plane as
\bea\label{eq:pmagnetic}
B^{\rm s}(\bR,\bK)\equiv\ve_{\gamma\beta}\partial_{R_\gamma}[\mathcal{A}^D_\beta(\bR,\bK)+\mathcal{A}^{H,\parallel}_\beta(\bR,\bK)+\mathcal{A}^{H,\perp}_\beta(\bR,\bK)],
\eea
where $\ve_{\gamma\beta}$ is the 2D Levi-Civita symbol (not to be mistaken with the similar-looking symbol of the strain tensor $\e_{\alpha\beta}$). The pseudo magnetic field from the first term $\boldsymbol{\mathcal A}^D$ requires computing $\ve_{\gamma\beta}\partial_{R_\gamma} u_{\alpha\beta}(\bR)=\ve_{\gamma\beta}\partial_{R_\gamma}\partial_{R_\beta} \mathcal D_{\alpha}(\bR)=0$. That is, the displacement correction from strain acts as a vector-potential with no curl and hence cannot produce a pseudo magnetic field. This correction serves as a gauge transformation in the minimal coupling prescription \cite{de_juan_gauge_2013}. On the other hand, the pseudo magnetic field from the hopping induced terms (both $\mathcal{A}^{H,\parallel}$ and $\mathcal{A}^{H,\perp}$) require calculating $\ve_{\gamma\beta}\partial_{R_\gamma} \e_{\alpha\beta}(\bR)\neq0$ in general, leading to mimicking the effect of a magnetic field. 

Further, just like in electromagnetism, we can also define a pseudo electric field from this vector-potential:
\bea\label{eq:pElectric}
E^{\rm s}_\alpha(\bR,\bK)\equiv-\partial_t[\mathcal{A}^D_\alpha(\bR,\bK)+\mathcal{A}^{H,\parallel}_\alpha(\bR,\bK)+\mathcal{A}^{H,\perp}_\alpha(\bR,\bK)].
\eea
These fields switch sign between the pair of singular points at $\bK$ and $-\bK$ (an hence the prefix of pseudo). In addition to these, we can also define fields from $\varphi_r$ in different irreps as $$E^{\rm s}_{r,\alpha}(\bR,\bK)\equiv-\partial_{R_\alpha}\varphi_r(\bR,\bK).$$ The $r=0$ field would correspond to an electric field that doesn't actually flip sign between the singular point pairs and can serve as a real electric field.

\paragraph*{Strain induced shift of high-symmetry points and singular points:} Strain induced vector-potentials couple to the Hamiltonian in the form $\hat v_\alpha (q_\alpha+e\mathcal A_\alpha)$. Such a coupling shifts the singular point from $\bK$ point to ${\bK}-e\boldsymbol{\mathcal A}$. This shift comes from three sources: the lattice displacement ($\mathcal A^D$), changes to the in-plane, and that to out-of-plane hoppings($\mathcal A^{H,\parallel}$ and $\mathcal A^{H,\perp}$). Each has a different consequence. The displacement shift is associated with the deformation of the BZ vectors from the unstrained case: $-e\mathcal A^D_\beta=-K_{\alpha}u_{\alpha\beta}$ (see the introduction to the Appendix). This is the shift that is not associated with the pseudo magnetic field and is also independent of the details of the unit cell. The other two shifts which arise from hopping corrections additionally move the singular point. As a consequence, if a singular point in the unstrained case were also a high-symmetry point of the BZ, then in the strained case, the shift due to the displacement part would not decouple the singular point from the high-symmetry point. The hopping induced shifts, on the other hand, would. In the context of graphene, this has been discussed in the literature under the label of `frame-effects' \cite{de_juan_gauge_2013}.

\section{Applications and extensions}\label{Sec:Examples}
We now apply our results to commonly used toy models of square and triangular lattices with 1, 2 and 3 atoms per unit cell. These models are not restricted to materials and are also useful for photonic lattices, systems of coupled resonators, or any system that could be mapped to a tunable tight-binding model. The second part of this section will deal with an example of modelling strain in a system with a high dimensional Hilbert space and reducing it to smaller dimensional subsystems. Henceforth, we will be suppressing the $\mathbf R$-dependence of the $\epsilon_{\alpha\beta}$ terms.

\begin{figure}[ht]
    \centering
    \includegraphics[width=0.9\textwidth]{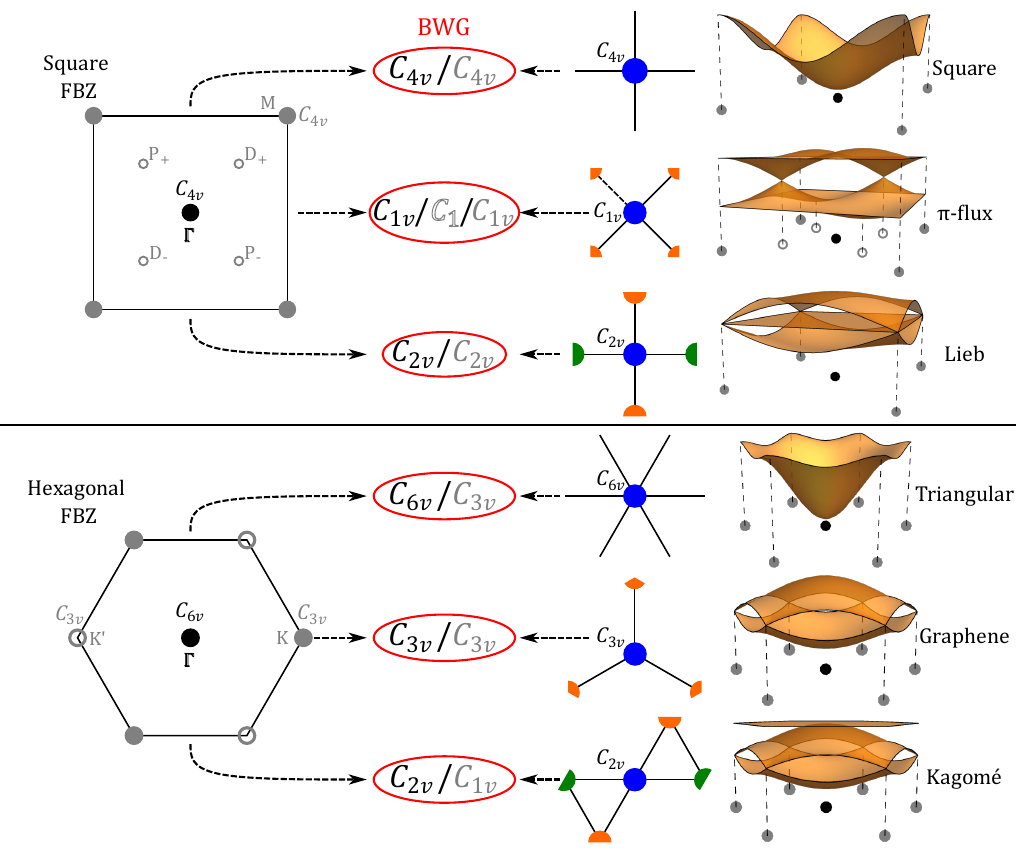}
    \caption{\textbf{The bond-wavevector groups}. The top panel shows the BZ of the square lattice along with the singular points corresponding to various scenarios discussed in the text. The corresponding wavevector groups are also stated. On the right are the bond geometries and the corresponding bond symmetry group for scenarios with 1,2 and 3 atoms per unit cell (square, $\pi$-flux, and Lieb). The dashed bond in $\pi$-flux system denotes a bond with a phase $e^{i\pi}$. The different colors represent the different atoms of the unit cell. The partial circles denote the corresponding contribution of the atom to the unit cell. Their band structures highlighting the extrema or points of degeneracies, i.e. the singular points, are also shown. The groups encircled in red represent the bond-wavevector group (BWG). The black/grey/empty color code of the groups correspond to those of the singular points in the BZ. The bottom row shows the same but for the triangular lattice systems: triangular, honeycomb(graphene), and Kagom\'e.}\label{Fig:3}
\end{figure}

\subsection{Strain induced potentials in 2D Bravais and non-Bravais systems}
In this subsection we consider strictly 2D models and write down the strain corrections and the corresponding potentials. Since there will be no out-of-plane bonds, the potentials $\varphi^{H,\perp},\boldsymbol{\mathcal{A}}^{H,\perp}$ would be absent. The supporting explicit calculations are presented in the Appendices. In all the cases we discuss below, $a$ will denote the lattice constant.

\subsubsection{Lattices with 1 atom per unit cell}\label{SubSec:1apuc}
Let us start with the very common system of a square lattice with nn hopping. This system has two relevant singular points, $\Gamma$-point (0,0) and the $M$-points $\{(\pm\pi,\pm\pi)/a\}$, where we can apply our results. Our first result allows us to write the strain corrected Hamiltonian in terms of the inverse mass tensors at those points, which are $\hat I_{xx}(\bK_{\Gamma/M})=\hat I_{yy}(\bK_{\Gamma/M})=\pm2a^2t$. This leads to [up to $\mathcal{O}(q^2)$] 
\bea\label{eq:Sq1GM}
h^{\rm Sq}(\bK_{\Gamma/M}+\bq)&=&\mp4t\underbrace{\pm2t\zeta\left(\e_{xx}+\e_{yy}\right)}_{s^H_\parallel}.
\eea
Here, we have used the fact that nn distance $\delta_{ab}$ is just the lattice constant $a$. These forms are corroborated by an explicit calculation in Appendix \ref{AppA}. Our second result allows us to interpret this correction as a potential in the irreps of the bond-wavevector group of the singular points. The bonds connecting a given atom to its nn have a $C_{4v}$ symmetry (see Fig. \ref{Fig:3}) and both the singular points in the BZ are also $C_{4v}$, making all the bond-wavevector groups also $C_{4v}$. The quadratic forms of this group are such that $x^2+y^2$ belongs to the scalar irrep $A_1$  (see Table \ref{Groups_table}), while $x^2-y^2,xy$ belong to irreps $B_1,B_2$. Since this is a 1 atom per unit cell system, the Hilbert space matrix structure is trivial with $\hat M^0(\bK_{\Gamma/M})=\mathds{1}_{1\times 1}$ being the only generator of the Hilbert space. As this is already being used to represent the scalar irrep, we cannot represent any other irreps. This is also why there are no velocity (vector) terms at these singular points as they would have needed generators of the $E$-irrep to couple with. Thus, we should expect the quadratic form to only show up as $x^2+y^2$ which leads to the form $\e_{xx}+\e_{yy}$. Indeed, that is what is seen in Eq. (\ref{eq:Sq1GM}). Next, following Eq. (\ref{AFormPhir}) with $r=0$, we find the scalar-potential to be
\bea\label{eq:SqscalarPot}
-e\varphi^H_0(\bK_{\Gamma/M})&=&\pm2t\zeta\left(\e_{xx}+\e_{yy}\right).
\eea

Another 1 atom per unit cell system would be the triangular lattice. Here there are 3 singular points: the $\Gamma$ and the $K,K'$ points of the hexagonal BZ (the coordinates of these  in our frame are listed in the Appendix). The $K,K'$ points form a $\bK,-\bK$ pair. The inverse masses are $\hat I_{xx}(\bK_{\Gamma})=\hat I_{yy}(\bK_{\Gamma})=3a^2t\hat M^0(\bK_\Gamma)$ and $\hat I_{xx}(\bK_{K/K'})=\hat I_{yy}(\bK_{K/K'})=-(3a^2t/2)\hat M^0(\bK_{K/K'})$, where $\hat M^0(\bK_{\Gamma/K/K'})=\mathds{1}_{1\times 1}$. Our first result allows us to write:
\bea\label{eq:T1GHams}
h^{\rm Tr}(\bK_{\Gamma}+\bq)&=&-6t\underbrace{+3t\zeta(\e_{xx}+\e_{yy})}_{s^H_\parallel},\nn\\
h^{\rm Tr}(\bK_{K/K'}+\bq)&=&3t\underbrace{-\frac32t\zeta(\e_{xx}+\e_{yy})}_{s^H_\parallel}.
\eea
We have used the fact that the nn distance is $a$. These are corroborated again with an explicit calculation in Appendix \ref{AppA}. To interpret these corrections as potentials, observe that the bonds here have a $C_{6v}$ symmetry (see Fig. \ref{Fig:3}), and the $\Gamma$ and the $K,K'$ points have $C_{6v}$ and $C_{3v}$ symmetry, respectively, making the bond-wavevector group to be $C_{6v}$ at the $\Gamma$ point and $C_{3v}$ at the $K,K'$ points. Once again, because the Hilbert space structure is trivial, we can only have scalar generators. This means that no velocity (vector) terms can appear and only the quadratic form $x^2+y^2$ should show up (Table \ref{Groups_table}). Following Eq. (\ref{AFormPhir}) with $r=0$, we can identify the scalar-potentials at these points as
\bea\label{eq:T1Kpots}
-e\varphi^H_0(\bK_\Gamma)&=&3t\zeta(\e_{xx}+\e_{yy}),\nn\\
-e\varphi^H_0(\bK_{K/K'})&=&-\frac32t\zeta(\e_{xx}+\e_{yy}).
\eea
We see that for systems with just one atom per unit cell, strain only acts as a grain dependent scalar correction which can also be seen as the modulation of the chemical potential.

\subsubsection{Non-Bravais lattices with 2 atoms per unit cell}\label{SubSec:2apuc}
For a square lattice, most 2-atoms per unit cell structures also exhibit zero velocity singular points and hence are similar to the 1-atom case. To really test our theory, consider the interesting scenario of the Hofstadter model of a square lattice in a magnetic field\cite{Hofstadter1976} at a strength that leads to $\pi$-flux linked per square plaquette. This lattice, in the particular gauge choice shown in Fig. \ref{Fig:3} has $C_{1v}$ symmetry. The full Hamiltonian is presented in Appendix \ref{AppB}. Due to the $\pi$-flux linkage the singular points are shifted from the high-symmetry points of the lattice at $\Gamma$ and $M$ to the interior points at $\pm(\pi,\pi)/2a$ [which have a Dirac (D) spectrum] and at $\pm(-\pi,\pi)/2a$ [which have a Parabolic (P) spectrum]. At the $D$-points, the relevant band structure parameters are
\bea\label{eq:PiD}
    \hat M^0(\bK_D^\pm)&=&0,\nn\\
    \hat v_{x}(\bK_D^\pm) &=& \pm at(\hat\sigma_x\mp\hat\sigma_y),\nn\\
    \hat v_{y}(\bK_D^\pm) &=& \pm at(\hat\sigma_x\pm\hat\sigma_y),\nn\\
    \hat I_{\alpha\beta}(\bK_D^\pm)&=&0~~\forall~\alpha,\beta.
\eea
From our first result, the strained Hamiltonian can be written as
\bea\label{eq:mid}\hat h^\pi(\bK_D^\pm+\bq)&=&\hat v_{\alpha}(aq_\alpha+ae\mathcal A^D_\alpha),\nn\\
e\mathcal A^D_\alpha(\bK_D^\pm)&=&K^\pm_{D,\beta} u_{\beta\alpha}.\eea 
In this case, the displacement induced vector-potential exists because of finite velocity terms [see Eq. (\ref{AFormAD})]. At the $P$-points, the band structure parameters are
\bea\label{eq:PiP}
    \hat M^{0}(\bK_P^\pm)&=&\frac{\hat\sigma_x\pm\hat\sigma_y}{\sqrt{2}}, \nn\\
    \hat v_{\alpha}(\bK_P^\pm) &=& 0~~\forall~\alpha,\nn\\
    \hat I_{xx}(\bK_P^\pm)&=& \hat I_{yy}(\bK_{P^\pm}) = (a^2t/\sqrt 2)\hat M^{0}(\bK_P^\pm),\nn\\
    \hat I_{xy}(\bK_P^\pm) &=& (a^2t/\sqrt 2)\hat M^{1}(\bK_P^\pm),\nn\\
    \text{where, } \hat M^{1}(\bK_P^\pm)&=&\frac{\hat\sigma_x\mp\hat\sigma_y}{\sqrt{2}}.
\eea
Our first result allows us to write
\bea\label{eq:stHAM}\hat h^\pi(\bK^\pm_P+\bq)&=&-2\sqrt 2t\hat M^{0}(\bK^\pm_P)+\underbrace{\sqrt 2t\zeta(\e_{xx}+\e_{yy})\hat M^{0}(\bK^\pm_P)+ 2\sqrt 2t\zeta\e_{xy}\hat M^{1}(\bK^\pm_P)}_{\hat s^H_\parallel},
\eea 
where we have used the fact that the nn distance is $a/\sqrt{2}$. These forms are corroborated in Appendix \ref{AppB} with an explicit calculation. 

To apply our second result, we note that we don't have anything to project at the $D$-points as they do not have a $s^H_\parallel$ correction. For the $P$-points, whose wavevector groups are $C_{1}$, the bond-wavevector groups are also $C_1$ (see Fig. \ref{Fig:3}). In this trivial group the linear and quadratic forms are all scalars. In this lattice, however, there are special emergent symmetries due to the $\pi$-flux that prevents the mixing of the odd and even momentum terms and also guarantees the presence of Dirac points in the spectrum \cite{Wen1989}\footnote{This is a different Dirac point from the one we are used to in graphene as the Hilbert space transforms differently under time reversal in the two cases.}. We won't discuss these here, but in the context of our results, the separation of the odd (linear) and even (quadratic) terms implies that either the velocity (linear) is finite and $s^H$ (quadratic) term is absent, or vice versa. The former happens at the $D$-points while the latter at the $P$-points. Since $C_1$ only has a scalar irrep, all the three quadratic forms belong to it (see Table \ref{Groups_table}). Thus, one would expect up to three generators in this irrep. However, since the Hilbert space dimension is 2, and this lattice is bipartite, we can only have two generators (composed of $\hat\sigma_x$, $\hat\sigma_y$). These generators can be expressed as $\hat M^0$ and $\hat M^1$ defined above. Following Eq. (\ref{AFormPhir}) with $r=0,1$ (since both are scalars), we find the two scalar-potentials in this case to be:
\bea
-e\varphi^H_0(\bK_P^\pm)&=&\sqrt 2t\zeta(\e_{xx}+\e_{yy}),\nn\\
-e\varphi^H_1(\bK_P^\pm)&=&\sqrt 2t\zeta(2\e_{xy}).
\eea

For the triangular lattice, let us consider the celebrated model of graphene with nn bonds which have the symmetries of $C_{3v}$ (Fig. \ref{Fig:3}). Here, the singular points are at the $\Gamma$ and the $K/K'$ points of the BZ. The full Hamiltonian is presented in Appendix \ref{AppB}. The relevant band structure parameters at the $\Gamma$-point are
\bea\label{eq:GrG}
    \hat M^0(\bK_\Gamma)&=&\hat \sigma_x,\nn\\
    \hat v_{\alpha}(\bK_\Gamma) &=& 0~~\forall~\alpha,\nn\\
    \hat I_{xx}(\bK_\Gamma)&=& \hat I_{yy}(\bK_\Gamma) = (a^2t/2)\hat \sigma_x,\nn\\
    \hat I_{xy}(\bK_\Gamma)&=&0.
\eea
Accounting for the nn bond distance of $a/\sqrt{3}$, our first result then implies that
\bea\label{eq:GrGHam}
\hat h^{\rm Gr}(\bK_\Gamma+\bq)&=&-3t\hat M^0(\bK_\Gamma)+\underbrace{\frac32t\zeta\left(\e_{xx}+\e_{yy}\right)\hat M^0(\bK_\Gamma)}_{\hat s^H_\parallel}.
\eea 
To apply the second result, note that the $\Gamma$-point has a C$_{6v}$ symmetry, making the bond-wavevector group $C_{3v}$. This group has quadratic forms in the scalar irrep $A_1$ and the vector irrep $E$ (see Table \ref{Groups_table}). However, since $\hat \sigma_x$ already represents the scalar matrix, we no longer have a pair of off-diagonal matrices in the $2\times2$ Hilbert space to represent the vectors. Thus, the Hamiltonian can only be composed of scalar generators of the form $x^2+y^2$ and we cannot have velocity components. This is confirmed by Eq. (\ref{eq:GrG}). The scalar-potential can then be extracted from Eq. (\ref{AFormPhir}) with $r=0$ as:\bea\label{eq:GrGscalar}
-e\varphi^H_0(\bK_\Gamma)&=&\frac32t\zeta\left(\e_{xx}+\e_{yy}\right).
\eea
At the $K$-point, the band structure parameters are
\bea\label{eq:GrK}
    \hat M^0(\bK_{K})&=&0,\nn\\
    \hat v_{\alpha}(\bK_{K}) &=& -(\sqrt{3}at/2)\hat\sigma_\alpha~~\forall~\alpha,\nn\\
    \hat I_{xx}(\bK_{K})&=& -\hat I_{yy}(\bK_K) = -(a^2t/4)\hat\sigma_x,\nn\\
    \hat I_{xy}(\bK_{K})&=&(a^2t/4)\hat\sigma_y.
\eea
This leads to the strained Hamiltonian being:
\bea\label{eq:GrKHam}
\hat h^{\rm Gr}(\bK_K+\bq)&=&-\frac{\sqrt{3}t}{2}[\hat{\sigma}_x(aq_x+ae\mathcal A^D_x)+\hat\sigma_y(aq^y+ae\mathcal A^D_y)]\underbrace{-\frac34t\zeta[(\e_{xx}-\e_{yy})\hat\sigma_x-2\e_{xy}\hat\sigma_y]}_{\hat s^H_\parallel},\nn\\
e\mathcal A^D_\alpha(\bK_K)&=&K_{K,\beta}u_{\beta\alpha}.
\eea 
Here we have a finite displacement-induced vector-potential due to the presence of finite velocity in the Hamiltonian [Eq. (\ref{AFormAD})]. To apply our second result, observe that the bond-wavevector group is still $C_{3v}$, like at the $\Gamma$-point, and thus has the quadratic forms in the scalar irrep $A_1$ and the vector irrep $E$. However, since the scalar term is zero at this singular point, we have both generators $\hat\sigma_x$ and $\hat\sigma_y$ available to represent the E irrep. Thus, we should expect the quadratic forms $x^2-y^2$ and $xy$ to couple as a vector in the Hamiltonian. Indeed, using Eq. (\ref{AFormPhir}) we confirm that there are no scalar-potentials and using Eq. (\ref{AFormAH}) we find the vector-potentials to be
\bea\label{eq:GrKvecpots}
ae\mathcal A^H_x(\bK_K)&=&\frac{\sqrt{3}}{2}\zeta(\e_{xx}-\e_{yy}),\nn\\
ae\mathcal A^H_y(\bK_K)&=&-\sqrt{3}\zeta\e_{xy}.
\eea

While there are no scalar corrections at the $K$-point in the nn model, if one were to consider the next nn hopping ($t'$), this would add another order of bonds that would essentially mimic the 1 atom per unit cell results of the triangular lattice resulting in a scalar-potential of the form in Eq. (\ref{eq:GrGscalar}) with $t\rightarrow t'$ \cite{McRae2019, wang_global_2021}.

\subsubsection{Non-Bravais lattices with 3 atoms per unit cell}\label{SubSec:3apuc}
For a square lattice, an example here of a system with an interesting band structure is the Lieb lattice \cite{Lieb1989}. It has a flat band and a rare triple degeneracy singular point at the $M$ points (see Fig. \ref{Fig:3}). The bonds here are $C_{2v}$ and the full Hamiltonian is presented in Appendix \ref{AppC}. For the singular point at $\Gamma$, the band structure parameters are
\bea\label{eq:LiG12}
     \hat M^{0}(\bK_{\Gamma})&=&\frac{\sqrt 3}{2}\begin{pmatrix}
        0&1&1\\
        1&0&0\\
        1&0&0
    \end{pmatrix},\nn\\
    \hat v_{\alpha}(\bK_{\Gamma}) &=& 0~\alpha,\nn\\
    \hat I_{xx}(\bK_{\Gamma})&=&(a^2t/2\sqrt{3})[\hat M^0(\bK_\Gamma)+\hat M^1(\bK_\Gamma)],\nn\\
    \hat I_{yy}(\bK_\Gamma)&=&(a^2t/2\sqrt{3})[\hat M^0(\bK_\Gamma)-\hat M^1(\bK_\Gamma)],\nn\\
    \hat I_{xy}(\bK_{\Gamma})&=&0,\nn\\
    \text{where}~\hat M^{1}(\bK_{\Gamma})&=&\frac{\sqrt 3}{2}\begin{pmatrix}
        0&1&-1\\
        1&0&0\\
        -1&0&0
    \end{pmatrix}.
\eea
Using the fact that the nearest neighbor bond length is $a/2$, we invoke our first result to write down the strained Hamiltonian as
\bea\label{eq:LiGHam}
\hat h^{\rm Li}(\bK_\Gamma+\bq)&=&-(4t/\sqrt{3})\hat M^{0}(\bK_\Gamma)+\underbrace{(2t\zeta/\sqrt{3})[\e_{xx}\{\hat M^0(\bK_\Gamma)+\hat M^1(\bK_\Gamma)\}+\e_{yy}\{\hat M^0(\bK_\Gamma)-\hat M^1(\bK_\Gamma)\}]}_{s^H_\parallel}.
\eea 
This form is corroborated with the explicit calculation in Appendix \ref{AppC}. To apply the second result, we note that the bond-wavevector group at $\Gamma$-point is $C_{2v}$ (see Fig. \ref{Fig:3}). In this group, the quadratic forms $x^2+y^2, x^2-y^2$ belong to $A_1$, whereas $xy$ belongs to $A_2$ (see Table \ref{Groups_table}). The vector irreps containing the linear terms $x,y$ are $B_1,B_2$. Because the scalars are already represented in the non-zero entries of the Hilbert space (and there are two of them), we don't have any generators left to represent $B_1,B_2$ and thus there cannot be any other irreps. Thus, we cannot have any vector-potential at the $\Gamma$-point, but following Eq. (\ref{AFormPhir}), the two potentials with $r=0,1$ can be extracted as:
\bea\label{eq:liebpots}
-e\varphi^H_0(\bK_\Gamma)&=&\frac{2t\zeta}{\sqrt 3}(\e_{xx}+\e_{yy}),\nn\\
-e\varphi^H_1(\bK_\Gamma)&=&\frac{2t\zeta}{\sqrt 3}(\e_{xx}-\e_{yy}).
\eea
Since $C_{2v}$ has both $x^2+y^2$ and $x^2-y^2$ under $A_1$ irrep, both $r=0$ and $r=1$ contributions would serve as scalar-potentials.

The band structure parameters at the $M$-point are
\bea\label{eq:LiM12}
     \hat M^0(\bK_{M})&=&0,\nn\\
    \hat v_{x}(\bK_{M}) &=& at\begin{pmatrix}
        0&1&0\\
        1&0&0\\
        0&0&0
    \end{pmatrix},~
    \hat v_{y}(\bK_{M}) = at\begin{pmatrix}
        0&0&1\\
        0&0&0\\
        1&0&0
    \end{pmatrix},\nn\\
    \hat I_{\alpha\beta}(\bK_{M})&=&0~~\forall~\alpha,\beta.
\eea
Collecting all these, the strained Hamiltonian can be written as
\bea\label{eq:LiMHam}
\hat h^{\rm Li}(\bK_M+\bq)&=&\hat v_{x}(aq_x+ae\mathcal A^D_x)+\hat v_{y}(aq_y+ae\mathcal A^D_y),\nn\\
e\mathcal A^D_\alpha(\bK_{M})&=&K_{M,\beta}u_{\beta\alpha}.
\eea 
At the $M$-point, the bond-wavevector group is still $C_{2v}$. However, because the scalar term is absent, the two generators can be used to represent the vector irreps $B_1,B_2$. This is why the velocity terms are present and so are the displacement induced potentials. But since the quadratic forms belong to non-vector irreps that are no longer possible to be represented, we do not have any $\hat s^H_\parallel$ corrections, and hence no additional potentials.

A relevant example for the triangular lattice in this category would be the Kagom\'e lattice which has become popular in the context of the study of correlated systems \cite{Ghimire2020}. Here, there are 3 bonds which are each $C_{2v}$ (Fig. \ref{Fig:3}), with the singular points at $\Gamma$ and $K/K'$. The full Hamiltonian is presented in Appendix \ref{AppC}. The band structure parameters at the $\Gamma$-point are:
\bea\label{eq:Kagscalar}
\hat M^0(\bK_\Gamma)&=&\frac1{\sqrt{2}}\begin{pmatrix}
    0&1&1\\
    1&0&1\\
    1&1&0
\end{pmatrix},\nn\\
\hat v_{\alpha}(\bK_\Gamma)&=&0~~\forall~\alpha,\nn\\
\hat I_{xx}(\bK_\Gamma)&=&\frac{a^2t}{4}[\sqrt{2}\hat M^0(\bK_\Gamma)-\hat M^1(\bK_\Gamma)],\nn\\
\hat I_{yy}(\bK_\Gamma)&=&\frac{a^2t}{4}[\sqrt{2}\hat M^0(\bK_\Gamma)+\hat M^1(\bK_\Gamma)],\nn\\
\hat I_{xy}(\bK_\Gamma)&=&\frac{a^2t}{4}\hat M^2(\bK_\Gamma),\nn\\
\text{where}~\hat M^1(\bK_\Gamma)&=&\frac12\begin{pmatrix}
    0&1&-2\\
    1&0&1\\
    -2&1&0
\end{pmatrix},~\hat M^2(\bK_\Gamma)=\frac{\sqrt 3}{2}\begin{pmatrix}
    0&1&0\\
    1&0&-1\\
    0&-1&0
\end{pmatrix}.
\eea
Using the fact that the nn bond distance is $a/2$, our first result implies that the strained Hamiltonian should be
\bea\label{eq:HamGKag}
\hat h^{\rm Kag}(\bK_\Gamma)&=&-2\sqrt{2}t\hat M^0(\bK_\Gamma)\underbrace{+t\zeta[\e_{xx}\{\sqrt{2}\hat M^0(\bK_\Gamma)-\hat M^1(\bK_\Gamma)\}+\e_{yy}\{\sqrt{2}\hat M^0(\bK_\Gamma)+\hat M^1(\bK_\Gamma)\}+2\e_{xy}\hat M^2(\bK_\Gamma)]}_{\hat s^H_\parallel}.
\eea
For the second result, we note that the bond-wavevector group at the $\Gamma$ point is $C_{2v}$ (see Fig. \ref{Fig:3}). The three quadratic terms, and hence their generators, already fall under the $A_1$ and $A_2$ irreps (see Table \ref{Groups_table}), leaving none for the vector irreps. Hence, there cannot be a velocity term. The potentials corresponding to these generators from Eq. (\ref{AFormPhir}) with $r=0,1,2$ are:
\bea\label{eq:kaggamscalar}
-e\varphi^H_0(\bK_\Gamma)&=&\sqrt 2t\zeta(\e_{xx}+\e_{yy}),\nn\\
-e\varphi^H_1(\bK_\Gamma)&=&-t\zeta(\e_{xx}-\e_{yy}),\nn\\
-e\varphi^H_2(\bK_\Gamma)&=&2t\zeta\e_{xy}.
\eea
While $r=0,1$ serve as scalar-potentials, the $r=2$ potential that falls under the $A_2$ irrep would serve as a `pseudo-scalar' potential. 

The band structure parameters at the $K$-point are:
\bea\label{eq:Kagvector}
\hat M^0(\bK_K)&=&\frac1{\sqrt 2}\begin{pmatrix}
    0&1&-1\\
    1&0&1\\
    -1&1&0
\end{pmatrix},\nn\\
\hat v_x(\bK_K)&=&-\frac{\sqrt{3}at}{2}\hat M^1(\bK_K),~\hat v_y(\bK_K)=-\frac{\sqrt{3}at}{2}\hat M^2(\bK_K),\nn\\
\hat I_{xx}(\bK_K)&=&\frac{a^2t}8\left[\sqrt{2}\hat M^0(\bK_K)-\hat M^1(\bK_K)\right],\nn\\
\hat I_{yy}(\bK_K)&=&\frac{a^2t}8\left[\sqrt{2}\hat M^0(\bK_K)+\hat M^1(\bK_K)\right],\nn\\
\hat I_{xy}(\bK_K)&=&\frac{a^2t}8\hat M^2(\bK_K).\nn\\
\text{where, }\hat M^1(\bK_K)&=&\frac{1}{2}\begin{pmatrix}
    0&1&2\\
    1&0&1\\
    2&1&0
\end{pmatrix},~\hat M^2(\bK_K)=\frac{\sqrt{3}}{2}\begin{pmatrix}
    0&1&0\\
    1&0&-1\\
    0&-1&0
\end{pmatrix},
\eea
Applying our first result, the strained Hamiltonian is
\bea\label{eq:h12Kag}
\hat h^{\rm Kag}(\bK_K+\bq)&=&-\sqrt{2}t\hat M^0(\bK_K)+\hat v_x(\bK_K)(aq_x+ae\mathcal A^D_x)+\hat v_y(\bK_K)(aq_y+ae\mathcal A^D_y)+\nn\\
&&~~~~~\underbrace{\frac{t\zeta}{2}[\e_{xx}\{\sqrt{2}\hat M^0(\bK_K)-\hat M^1(\bK_K)\}+\e_{yy}\{\sqrt{2}\hat M^0(\bK_K)+\hat M^1(\bK_K)\}+2\e_{xy}\hat M^2(\bK_K)]}_{\hat s^H_\parallel},\nn\\
\text{where, }e\mathcal A^D_\alpha(\bK_K)&=&K_{K,\beta}u_{\beta\alpha}.
\eea
This is corroborated by an explicit calculation in Appendix \ref{AppC}. To apply the second result, we note that the bond-wavevector group here is $C_{1v}$. The quadratic forms are in the irreps $A_1$ and $A_2$, which also contain the vectors. The Hilbert space here needs three generators, which are provided by $\hat M^{0/1/2}$ identified above. It is evident from Eq. (\ref{eq:Kagvector}) that $\hat M^{1/2}$ serve as vector generators. Using Eqs. (\ref{AFormPhir}) and (\ref{AFormAH}), the scalar- and vector-potentials are then identified as:
\bea\label{eq:regroup}
-e\varphi^H_0(\bK_K)&=&\frac{t\zeta}{\sqrt 2}(\e_{xx}+\e_{yy}),\nn\\
ae\mathcal A^H_x(\bK_K)&=& \frac{t\zeta}{\sqrt{3}}(\e_{xx}-\e_{yy}),\nn\\
ae\mathcal A^H_y(\bK_K)&=& -\frac{t\zeta}{\sqrt{3}}(2\e_{xy}).
\eea
Observe that the vector-potential from $\hat s^H_\parallel$ exists because the bond-wavevector group in $C_{1v}$. Observe also that the potential $\mathcal A^D_\alpha$ at the K point here is the same as in the 2 atom per unit cell graphene as this potential only depends on the lattice but not the details of the unit cell.

\subsection{Strain induced potentials from Hilbert space projections}\label{SubSec:Projections}
While we have demonstrated how one can use our two results to write down strain corrections and interpret them in terms of potentials is various irreps of the bond-wavevector group, we show here that effective strain-corrections and potentials can also be derived for a smaller Hilbert subspace of a larger system. This is done by projecting out the undesirable Hilbert space using L\"owdin's method \cite{lowdin_note_1951,Bae2023}. We also show that the rules we identified in this work with respect to the bond-wavevector group also apply to the new effective bonds formed in the projected subspace.

As an example, consider the super-Honeycomb lattice \cite{Aoki1996,Lan2012}. This system is bipartite with 5 atoms per unit cell with the two subsystems being $\{X,Y\}$ and $\{A,B,C\}$ (see Fig. \ref{Fig:4}). It's Hamiltonian is provided in Appendix \ref{AppD}. Although the lattice is $C_{6v}$, the bonds here are all $C_{1v}$ (Fig. \ref{Fig:4}) and hence all singular points ($\Gamma,K,K'$) also have the $C_{1v}$ bond-wavevector symmetry. Generally speaking, the $C_{1v}$ nature is going to be a common scenario in higher dimensional Hilbert spaces. Applying our first and second results, we can write the Hamiltonian at the $\Gamma$-point to be (we drop the arguments $\bK_\Gamma$ for brevity):
\bea\label{eq:H5Ham}
\hat h(\bK_\Gamma+\bq)&=&-\frac{t\sqrt{12}}{\sqrt{5}}\hat M^0+\frac{at}{\sqrt{10}}\hat M^x+\frac{at}{\sqrt{10}}\hat M^y+\underbrace{\frac{t\zeta\sqrt{3}}{\sqrt{10}}\left[\e_{xx}(\sqrt{2}\hat M^0+\hat M^2)+\e_{yy}(\sqrt{2}\hat M^0-\hat M^2)+2\e_{xy}\hat M^1\right]}_{\hat s^H_\parallel},\nn\\
\text{where, }\hat M^r&=&\begin{pmatrix}
    0_{2\times2}&\hat b_r\\\hat b^\dag_r&0_{3\times 3}
\end{pmatrix},\nn\\
\text{and } \hat b_0(\bK_\Gamma)&=&\sqrt{\frac5{12}}\begin{pmatrix}
    1&1&1\\
    1&1&1
\end{pmatrix},~\hat b_1(\bK_\Gamma)=\sqrt{\frac58}\begin{pmatrix}
    1&0&-1\\
    1&0&-1
\end{pmatrix},~\hat b_2(\bK_\Gamma)=\sqrt{\frac5{24}}\begin{pmatrix}
    1&-2&1\\
    1&-2&1
\end{pmatrix},\nn\\
\hat b_x(\bK_\Gamma)&=&\sqrt{\frac58}\begin{pmatrix}
    i&0&-i\\
    -i&0&i
\end{pmatrix},~\hat b_y(\bK_\Gamma)=\sqrt{\frac5{24}}\begin{pmatrix}
    i&-2i&i\\
    -i&2i&-i
\end{pmatrix}.
\eea
Note that although this is a $5\times5$ Hilbert space with $24$ generators plus identity, the result is expressed only in terms of the three scalar generators $\hat M^{0/1/2}$ and the two vector generators, $\hat M^{x/y}$. The various potentials from Eq. (\ref{AFormPhir}) are:
\bea\label{eq:H5scalar}
-e\varphi^H_0(\bK_\Gamma)&=&\sqrt{\frac{3}{5}}t\zeta(\e_{xx}+\e_{yy}),\nn\\
-e\varphi^H_1(\bK_\Gamma)&=&\sqrt{\frac{3}{10}}t\zeta(2\e_{xy}),\nn\\
-e\varphi^H_2(\bK_\Gamma)&=&\sqrt{\frac{3}{10}}t\zeta(\e_{xx}-\e_{yy}).
\eea
Since this uses up all the $\hat s_\parallel^H$ components, there are no vector-potentials in this case. Observe that the generators of the quadratic forms are different from the generators of the vectors even though they belong to the same irreps. This is a consequence of a large Hilbert space which permits many generators in a single irrep. The analysis is the same at the $K$-point where the matrix elements of the Hamiltonian only differ by a phase (see Appendix \ref{AppD}). The only difference would be that this case also includes the displacement induced vector-potential $\mathcal A^D_\alpha$.

\begin{figure}[ht]
    \centering
    \includegraphics[width=0.9\textwidth]{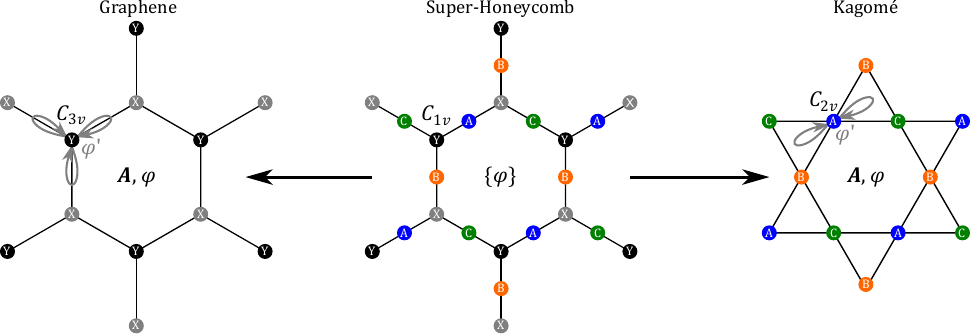}
    \caption{\textbf{The super-Honeycomb lattice and its Hilbert space projections}: This lattice has 
 two families of bonds $X\rightarrow A,B,C$ and $Y\rightarrow A,B,C$, each with a $C_{1v}$ symmetry. Strain creates a set of three scalar-potentials. Upon projection, we retrieve the graphene/Kagom\'e bonds and also their associated potentials. We also get additional atom-specific scalar-potentials (the grey $\varphi'$) upon projection in each case.}\label{Fig:4}
\end{figure}

\paragraph*{Hilbert space projection to the $XY$ subspace:} The super-Honeycomb lattice has a special feature that one can project the 5-atom Hilbert space on to either the XY (honeycomb) subspace or the ABC (Kagom\'e) subspace (see Fig \ref{Fig:4}). These projections are carried out using L\"owdin's method \cite{lowdin_note_1951,Bae2023} at some reference energy $E_0$ \footnote{Formally the two subsystems should be separated in energy. This is easily done by introducing a unit matrix in one of the subspaces. But we ignore it here for simplicity.}. Because this system is bipartite, the projection procedure is rather simple. Doing so on to honeycomb gives the Hamiltonian
\bea \label{eq:GrProjSM}
\hat H^{\rm Proj}_G(\bK_K+\bq)&=&\frac{\hat b\hat b^{\dag}}{E_0}\nn\\
&=&\frac{t^2}{E_0} \left(3\hat\sigma_0+\frac{\sqrt3}{2}\hat\sigma_x(aq_x+ae\mathcal A^D_x)+\frac{\sqrt3}{2}\hat\sigma_y(aq_y+ae\mathcal A^D_y)+\right.\nn\\
&&~~~~~~~~\left.\underbrace{-\zeta(\e_{xx}+\e_{yy})\hat\sigma_0+\frac{3\zeta}{2}\hat\sigma_x(\e_{xx}-\e_{yy})-3\zeta\hat\sigma_y\e_{xy}]}_{\hat s^H_\parallel}\right),\nn\\
\text{where, }e\mathcal A^D_\alpha&=& K_{K,\beta}u_{\beta \alpha}.
\eea
The matrix $\hat b$ is defined in Appendix \ref{AppD}. Up to a factor of 2, this has the same strain correction and vector-potential as in Eqs. (\ref{eq:GrKHam}) and (\ref{eq:GrKvecpots}). This can be attributed to the fact that after projection, the resulting bonds in the system (see Fig. \ref{Fig:4}) have a $C_{3v}$ symmetry and should thus respect the bond-wavevector based classification of the strain corrections. What is new in this case is that the projection also introduces a zeroth order (on-site) hopping $\propto(\e_{xx}+\e_{yy})$. In this order the Hilbert space is decoupled for the $X$ and $Y$ atoms. Thus, only scalar function would be permissible. The symmetry of the zeroth order bond is determined by that of the virtual bond with the atom in the subspace that has been projected out. In this case the bond symmetry would be $C_{3v}$ (see Fig. \ref{Fig:4}, Graphene projection), making the bond-wavevector group also $C_{3v}$. Since the only quadratic function in the scalar irrep is $x^2+y^2$, $\e_{xx}+\e_{yy}$ should be the only combination to appear for the zeroth order hoppings, which is indeed the case in Eq. (\ref{eq:GrProjSM}).

\paragraph*{Hilbert space projection to the $ABC$ subspace:} For the Kagom\'e projection we would get
\bea \label{eq:KagProjSM}
&&H^{\rm Proj}_K(\bK_K+\bq)=\frac{\hat b^{\dag}\hat b}{E_0}\nn\\
&&=\frac{t^2}{E_0}
\begin{pmatrix} 
    2+s_A & 1+s+\frac{\sqrt3}{4}ap_x+\frac{3}{4}ap_y& -1-s+\frac{\sqrt3}{2}ap_x\\
    1+s+\frac{\sqrt3}{4}ap_x+\frac{3}{4}ap_y & 2 +s_B& 1+s+\frac{\sqrt3}{4}(aq_x+aeA_x)-\frac{3}{4}ap_y\\
    -1-s+\frac{\sqrt{3}}{2}ap_x & 1+s+\frac{\sqrt3}{4}ap_x-\frac{3}{4}ap_y & 2+s_C
\end{pmatrix} 
\eea
where
\bea\label{eq:defsp}
p_\alpha&\equiv& q_\alpha+e\mathcal A_\alpha\nn\\
e\mathcal A_x&=&K_{K,\beta}u_{\beta x}+\frac{\zeta}{\sqrt{3}a}(\e_{xx}-\e_{yy})\nn\\
e\mathcal A_y&=&K_{K,\beta}u_{\beta y}+\frac{\zeta}{\sqrt{3}a}(-2\e_{xy})\nn\\
s&=&-\zeta(\e_{xx}+\e_{yy})\nn\\
s_A&=&-\zeta(3\epsilon_{xx}+\epsilon_{yy} + 2\sqrt3\epsilon_{xy})\nn\\
s_B&=&-\zeta(4\epsilon_{yy})\nn\\
s_C&=&-\zeta(3\epsilon_{xx}+\epsilon_{yy} - 2\sqrt3\epsilon_{xy}).
\eea
Once again, the hoppings match that of a Kagom\'e lattice and thus have the same bond-wavevector group. We should thus expect the scalar and vector-potentials of Eq. (\ref{eq:regroup}) to also show up here. Like in the case of projected Graphene, there are the zeroth order hoppings which arise from virtual hoppings to atoms that have been projected out. These bonds were $C_{1v}$ each and hence they correspond to the quadratic form that is along the distance of the bond. There are 3 such sites and hence there are 3 scalar-potentials $\propto s_A,s_B,s_C$.

It is worth noting that although the parent system did not have the required bond-symmetry for the presence of a vector-potential, it emerged in the projected system. The fact that our theory applies even to the projected system should not be surprising as it is based on group theory which only cares about the appropriate symmetry of the bonds in any system, whether or not it is derived from projection.

\section{Strained bilayer Graphene}\label{Sec:3D}
Consider the system of Bernal stacked bilayer graphene \cite{Yan2011}. This system has regained popularity due to a rich phase diagram of broken symmetry phases where superconductivity has been demonstrated \cite{Zhou2022} and shown to be boosted by proximity to a TMDC substrate \cite{Zhang2023}. Due to the presence of a substrate, strain could be expected to play a major role in altering the electronic correlations. We will use this multi-layer system to showcase how our rules can be used to calculate the strain correction to the Hamiltonian, and how different orders of bonds give rise to different potentials.

\begin{figure}[ht]
    \centering
    \includegraphics[width=\textwidth]{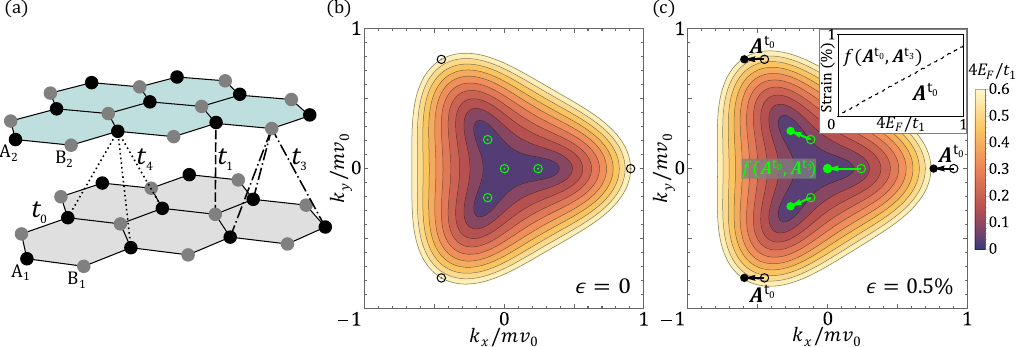}
    \caption{\textbf{Effect of multiple vector-potentials in bilayer graphene:} (a) Hoppings of different orders in bilayer graphene. Low energy spectrum of bilayer graphene under (b) no strain, (c) a uniaxial strain along the zigzag direction with $\e_{\alpha z}=0=\e_{xy}$, $\e_{xx}=\e=0.5\%$, and $\e_{yy}=-\nu \e$ (where $\nu$ is the Poisson ratio of bilayer graphene). The mass $m$ is defined in the text, $v_0=\sqrt{3}at_0/2$, and the hopping parameters used are those in \cite{kuzmenko_determination_2009,mccann_electronic_2013}. In (c) we see a shift of the corners of energy contours (empty $\rightarrow$ filled) due to strain. The shift at larger energies is captured already by $\mathcal A^{t_0}$, the potential from the intralayer hoppings of the monolayer. The shift of the Dirac cones (in green) at lower energies, however, depends also on $\mathcal A^{t_3}$ that is induced by interlayer hopping. The inset presents a separation of regimes where one would need to account for one vs multiple vector-potentials to correctly capture the electronic levels. The dashed line is a contour below which using only $\mathcal A^{t_0}$ leads to an error of less than 5$\%$ compared to accounting for both vector potentials.}\label{Fig:5}
\end{figure}

The Hamiltonian of bilayer graphene at the $K$ point, in the basis choice of $(A_1, B_1, A_2, B_2)$ (see Fig. \ref{Fig:5}a), where the subscript denotes the layer index, is \cite{mccann_electronic_2013}:
\bea\label{eq:BLG1}
    \hat H^0(\mathbf K_K +\mathbf q)&=& \begin{pmatrix}\hat h_{\rm intra} & \hat h_{\rm inter}\\
    \hat h^\dag_{\rm inter}& \hat h_{\rm intra}\end{pmatrix},\nn\\
\text{where }\hat h_{\rm intra}&=& -\frac{\sqrt 3}2\begin{pmatrix}0 & t_0aq_-\\
t_0aq_+&0\end{pmatrix} + \mathcal{O}(q^2),\nn\\
    \hat h_{\rm inter}&=&-\frac{\sqrt 3}2
    \begin{pmatrix}-t_4aq_- & t_3aq_+\\
    -\frac2{\sqrt 3}t_1&-t_4aq_-\end{pmatrix} + \mathcal{O}(q^2).
\eea
Here $q_{\pm}=q_x\pm iq_y$. We can identify four orders of bonds in this system. They are the intralayer hopping $t_0$, the directly vertical interlayer $AB$ hopping $t_1$, the nearest interlayer $AA$ hopping $t_{3}$, the second nearest interlayer $AB$ hopping $t_4$. Calculating the scalar, velocity and inverse mass matrices for each order, we get (listing only the non-zero entries):
\bea\label{eq:BLGparams}
t_0\text{-bonds}:~\hat v^{t_0}_\alpha&=&-\frac{\sqrt{3}at_0}{2}\begin{pmatrix}
    \hat\sigma_\alpha&0\\
    0&\hat\sigma_\alpha
\end{pmatrix},~\hat I_{xx}^{t_0}=-\hat I_{yy}^{t_0}=-\frac{a^2t_0}{4}\begin{pmatrix}
    \hat\sigma_x&0\\
    0&\hat\sigma_x
\end{pmatrix},~\hat I_{xy}^{t_0}=\frac{a^2t_0}{4}\begin{pmatrix}
    \hat\sigma_y&0\\
    0&\hat\sigma_y
\end{pmatrix};\nn\\
t_1\text{-bonds}:~\hat M^{t_1,0}&=&\begin{pmatrix}
    0&0&0&0\\
    0&0&1&0\\
    0&1&0&0\\
    0&0&0&0
\end{pmatrix};\nn\\
t_3\text{-bonds}: ~\hat v^{t_3}_{x}&=&-\frac{\sqrt{3}at_3}{2}\begin{pmatrix}
    0&0&0&1\\
    0&0&0&0\\
    0&0&0&0\\
    1&0&0&0
\end{pmatrix},~\hat v^{t_3}_{y}=-\frac{\sqrt{3}at_3}{2}\begin{pmatrix}
    0&0&0&i\\
    0&0&0&0\\
    0&0&0&0\\
    -i&0&0&0
\end{pmatrix},\nn\\
\hat I^{t_3}_{xx}&=&-\hat I^{t_3}_{yy}=-\frac{a^2t_3}{4}\begin{pmatrix}
    0&0&0&1\\
    0&0&0&0\\
    0&0&0&0\\
    1&0&0&0
\end{pmatrix},~\hat I^{t_3}_{xy}=\frac{a^2t_3}{4}\begin{pmatrix}
    0&0&0&i\\
    0&0&0&0\\
    0&0&0&0\\
    -i&0&0&0
\end{pmatrix};\nn\\
t_4\text{-bonds}:~\hat v^{\perp,t_4}_{x}&=&\frac{\sqrt{3}at_4}{2}\begin{pmatrix}
    0&0&1&0\\
    0&0&0&1\\
    1&0&0&0\\
    0&1&0&0
\end{pmatrix},~\hat v^{t_4}_{y}=\frac{\sqrt{3}at_4}{2}\begin{pmatrix}
    0&0&-i&0\\
    0&0&0&-i\\
    i&0&0&0\\
    0&i&0&0
\end{pmatrix},\nn\\
\hat I^{t_4}_{xx}&=&-\hat I^{t_4}_{yy}=\frac{a^2t_4}{4}\begin{pmatrix}
    0&0&1&0\\
    0&0&0&1\\
    1&0&0&0\\
    0&1&0&0
\end{pmatrix},~\hat I^{t_4}_{xy}=-\frac{a^2t_4}{4}\begin{pmatrix}
    0&0&-i&0\\
    0&0&0&-i\\
    i&0&0&0\\
    0&i&0&0
\end{pmatrix}.
\eea
Following Eqs. (\ref{eq:StrainClass}) and (\ref{eq:zstrain}), we can write down the strain correction from all orders to our Hamiltonian in terms of the above band structure parameters:
\bea\label{BLGst}
    \hat s^D &=& K_{K,\beta}u_{\beta \alpha}(\hat v^{t_0}_\alpha + \hat v^{t_3}_\alpha + \hat v^{t_4}_\alpha), \nn \\
    \hat s^H_\parallel &=&\begin{pmatrix}
    \hat s_{\rm intra}&\hat s_{\rm inter}\\
   \hat s^\dag_{\rm inter}&\hat s_{\rm intra}
\end{pmatrix},~~~\hat s^H_\perp =\begin{pmatrix}
    0&\hat z_{\rm inter}\\
   \hat z^\dag_{\rm inter}&0
\end{pmatrix},\nn\\
\text{where }\hat s_{\rm intra}&=&-\frac{3\zeta}{4}\begin{pmatrix}
    0&t_0(\e_{xx}-\e_{yy}+2i\e_{xy})\\
    t_0(\e_{xx}-\e_{yy}-2i\e_{xy})&0
\end{pmatrix},\nn\\
\hat s_{\rm inter}&=&-\frac{3\zeta a^2}{4(a^2+3b^2)}\begin{pmatrix}
    -t_4(\e_{xx}-\e_{yy}+2i\e_{xy})&t_3(\e_{xx}-\e_{yy}-2i\e_{xy})\\
    0&-t_4(\e_{xx}-\e_{yy}+2i\e_{xy})
\end{pmatrix}, \nn\\
\hat z_{\rm inter}&=&\frac{3\sqrt 3\zeta ab}{a^2+3b^2}\begin{pmatrix}
    -t_4(i\e_{xz}+\e_{yz})&t_3(i\e_{xz}-\e_{yz})\\
    -t_1\frac{a^2+3b^2}{3\sqrt 3\zeta ab}\e_{zz}&-t_4(i\e_{xz}+\e_{yz})
\end{pmatrix}.
\eea

We then use Eq. (\ref{Potentials}) to identify the following potentials:
\bea\label{eq:sectors}
t_0\text{-bonds}&:&e\mathcal A^{t_0}_x=K_{K,\beta}u_{\beta x}+\frac{\sqrt{3}\zeta}{2a}(\e_{xx}-\e_{yy}),\nn\\
&&e\mathcal A^{t_0}_y=\underbrace{K_{K,\beta}u_{\beta y}}_{\mathcal A^D}+\underbrace{\frac{\sqrt{3}\zeta}{2a}(-2\e_{xy})}_{\mathcal A^{0,\parallel}},\nn\\
t_1\text{-bonds}&:&e\varphi^\perp=t_1\zeta\e_{zz},\nn\\
t_3\text{-bonds}&:&e\mathcal A^{t_3}_x=K_{K,\beta}u_{\beta x}+\frac{\sqrt{3}\zeta a}{2(a^2+3b^2)}(\e_{xx}-\e_{yy})+\frac{6\zeta b}{a^2+3b^2}\e_{yz},\nn\\
&&e\mathcal A^{t_3}_y=\underbrace{K_{K,\beta}u_{\beta y}}_{\mathcal A^D}+\underbrace{\frac{\sqrt{3}\zeta a}{2(a^2+3b^2)}(-2\e_{xy})}_{\mathcal A^{3,\parallel}}\underbrace{-\frac{6\zeta b}{a^2+3b^2}\e_{xz}}_{\mathcal A^{3,\perp}},\nn\\
t_4\text{-bonds}&:&e\mathcal A^{t_4}_x=K_{K,\beta}u_{\beta x}+\frac{\sqrt{3}\zeta a}{2(a^2+3b^2)}(\e_{xx}-\e_{yy})-\frac{6\zeta b}{a^2+3b^2}\e_{yz},\nn\\
&&e\mathcal A^{t_4}_y=\underbrace{K_{K,\beta}u_{\beta y}}_{\mathcal A^D}+\underbrace{\frac{\sqrt{3}\zeta a}{2(a^2+3b^2)}(-2\e_{xy})}_{\mathcal A^{4,\parallel}}\underbrace{+\frac{6\zeta b}{a^2+3b^2}\e_{xz}}_{\mathcal A^{4,\perp}}.
\eea
The full strain corrected Hamiltonian is then obtained as:
\bea\label{eq:BLGstHam}
\hat h^{\rm BLG}(\bK_K+\bq)&=&(t_1-e\varphi^\perp)\hat M^0+\hat v^{t_0}_\alpha(aq_\alpha+ae\mathcal A^{t_0}_\alpha) + \hat v^{t_3}_\alpha(aq_\alpha+ae\mathcal A^{t_3}_\alpha)+\hat v^{t_4}_\alpha(aq_\alpha+ae\mathcal A^{t_4}_\alpha).
\eea

In general, potentials arising from different bonds will be independent. For example, in Eq. (\ref{eq:sectors}) we have three different vector-potentials for the orders identified by $t_0$, $t_3$ and $t_4$ bonds. However, geometrical constraints of a system can force these potentials to be the same. Indeed, if the out-of-plane shear $\e_{\alpha z}=0$, the potentials from $t_3$ and $t_4$ become the same due to these bonds having identical geometry of the $j$-neighbors, albeit rotated by $90^{\circ}$ (see Fig. \ref{Fig:5}a). In this case, we may even see that section of the system as having just one vector potential as opposed to two of them. The corresponding velocity projector would then be just be sum of the velocities of the two orders.

\subsection{Multiple vector-potentials}
Having established that every order of bonds can induce its own set of scalar- and vector-potentials, it is now pertinent to ask what impact it has in the analysis of the electronic structure. By extension, this will affect processes involving scattering, quantum tunnelling, etc., but this will not be addressed here. The first point to note would be that for lightly doped/gated systems, not all of 4 bands are relevant for low energy physics. Performing L\"owdin's projection on to the subspace of $(A_1,B_2)$ \cite{mccann_electronic_2013} we get the following low-energy effective Hamiltonian:
\bea\label{eq:BLG_Low}
    H^{\rm Low}_{\rm BLG}&=&\begin{pmatrix}0&-\frac{\sqrt{3}at_3}{2}(q_++e\mathcal A^{t_3}_+)-\frac{1}{2m}(q_++e \mathcal A^{t_0}_+)^2\\
    -\frac{\sqrt{3}at_3}{2}(q_-+e \mathcal A^{t_3}_-)-\frac{1}{2m}(q_-+e \mathcal A^{t_0}_-)^2&0\end{pmatrix},
\eea
with $m \equiv (1-\zeta\e_{zz})\frac{2t_1}{3a^2t_0^2}$ being the effective mass of the band, and $\mathcal A^p_\pm=\mathcal A^p_x\pm i\mathcal A^p_y$, for $p\in\{t_0,t_3\}$. Observe also that if $t_3\rightarrow0$, then the linear term would be absent. Had we started with the low energy model for the unstrained system, we would not have gotten the vector-potential coupling as the bands would have no velocity. This demonstrates the need of modelling strain in the appropriate Hilbert space and then projecting out the higher energy degrees of freedom.

The matrix elements are of the form $v(q+A^{t_3})+(q+A^{t_0})^2/2m =(q+A^{t_0}+mv)^2/2m + v(A^{t_3}-A^{t_0})-mv^2/2$. This form makes it clear that at high energies (large $q$) the system essentially acts like a parabolic system subject to the potential $A^{t_0}$, while at low energies (small q) the system is affected both by $A^{t_0}$ and the constant shift from $v(A^{t_3}-A^{t_0})$. This would notably affect the position of the Dirac points. Indeed this is what we find in Fig. \ref{Fig:5}b,c where we show the strain modification to the spectrum of bilayer graphene. The shift of the Dirac points at low energies is some non-linear function of both in- and out-of-plane vector-potentials $f(\mathbf{\mathcal A}^{t_0},\mathbf{\mathcal A}^{t_3})$ (here $\mathbf{\mathcal A}^{t_3}=\mathbf{\mathcal A}^{t_4}$ as $\e_{\alpha z}=0$). These are described in more detail in Refs. \cite{Mucha2011,Mariani2012}. However, the higher energies are only affected by a constant shift of $\mathbf{\mathcal A}^{t_0}$. This means that, for a given strain, there is a threshold energy (indicated by the dashed line in the Figure \ref{Fig:5}c inset) below which the system can be accurately modeled as behaving under the effects of a single vector-potential. This is particularly useful for analyzing experimental data (of gate dependent transport measurements, for example) where a sample would be appropriately doped and strain could be accounted for without the added complexity of multiple vector-potentials. Provided the strain and energy ranges are experimentally feasible like in bilayer graphene, this insight opens the door to a systematic treatment of strain in many other 2D materials, such as TMDCs.

\section{Conclusions}\label{Sec:Conclusions}
While it is known that strain corrections arise from two sources (displacements and change in hoppings as a result of displacements), in this work we have shown that the corrections from these two sources could actually be broken down into three parts and can be written directly in terms of the band structure parameters of the system. The displacement induced correction ($\hat s^D$) is the first part and couples to the velocity [Eq. (\ref{eq:StrainClass})]. The hopping induced term provides the other two parts as in-plane ($\hat s^H_\parallel$) and out-of-plane($\hat s^H_\perp$) contributions. The former is expressed in terms of the inverse mass tensor [Eq. (\ref{eq:StrainClass})] and the latter in terms of the velocity and the unstrained (scalar) form of the Hamiltonian [Eq. (\ref{eq:zstrain})]. In addition, we introduced the notion of orders of bonds that is classified based on distance and the type of bonds and showed that strain corrections can be written down for each order of the bonds.

Next, we introduced the notion of bond-wavevector group that takes into account the local symmetry of a given bond and allows us to regroup the inverse mass tensor into scalar and vector terms (in general, into all irreps of the bond-wavevector group). We identified Hilbert space generators that represent various scalars and vectors and prescribed a projection technique using these generators to extract scalar- and vector-potentials from the strain correction terms identified above [set of Eqs. (\ref{Potentials})]. We showed that a vector-potential like interpretation of in-plane strain is only possible for a system where the bond-wavevector group is $C_{1v}$ and $C_{3v}$. Since there are multiple orders of bonds, it is natural in our theory to expect multiple scalar- and vector-potentials.

We then demonstrated the applicability of our prescription in several toy models involving square and triangular lattices with multiple atoms per unit cell and showed that the popular system of graphene is a special case that respects all the conditions necessary to let strain serve as a vector-potential. We also demonstrated how one can start from a high dimensional strained Hilbert space and arrive at effective low energy models for strained Hamiltonians. We showed that the Hilbert space projection preserves the bond-wavevector based classification in the projected system. Finally, we applied our result to strained bilayer graphene and showed the existence of multiple vector-potentials, presented a Hilbert space projected low-energy model for strained bilayer graphene and identified energy regimes where one vector-potential would be sufficient and where one would need all the potentials to correctly capture the electronic structure. 

We believe that this general formulation opens up the door to many interesting investigations. It is easy to see that the approach is not limited to materials, but these principles extend to other engineered 2D systems. For example, strained carbon nanotubes of different chiralities can be modeled readily from this procedure. The effect of strain involving out-of-plane $\ve_{\alpha z}$ components becomes relevant in the case of hetero-strain \cite{Hou2024} where multiple layers are deformed differently. Any system that can be mapped to a tight-binding model, such as strained photonic systems, can also benefit from the principles outlined in this work. Time-varying strain and strain induced electric fields can be studied in systems beyond graphene for nano electro-mechanical systems based applications\cite{Low2012,Jiang2013,Xu2022}. The most immediate benefit in all of the above cases is that the theory readily provides a parametric model Hamiltonian that allows access to strain corrected wavefunctions which in turn allows us to model strain tuned ballistic quantum transport phenomena.

\paragraph*{Acknowledgments:} SM was funded by the Natural Sciences and Engineering Research Council of Canada (NSERC) Grant No. RGPIN-2019-05486. AC was funded by the Natural Sciences and Engineering Research Council of Canada (NSERC) Grant No. RGPIN-2019-06975. RZ was supported through a scholarship from the Fonds de Recherche du Québec - Nature et Technologies (FRQNT).

\appendix
\section*{APPENDIX}
In all of the following calculations, we follow the following convention. The translation vectors are $\bR_1,\bR_2$ which are of moduli $a$ each. We choose our frame such that for a square lattice, $\bR_1=(a,0),~\bR_2=(0,a)$, and for a triangular lattice, $\bR_1=(a,0),~\bR_2=(a/2,\sqrt 3 a/2)$. Finally, in all of the BZ vectors below, we shall absorb the lattice constant $a$ into the wavenumbers resulting in a dimensionless $\bk$. In this notation the BZ of the square lattice is marked with the $M$-points at $(\pm\pi,\pm\pi)$. The BZ of the triangular lattice is marked with the $K$-points at $(-4\pi/3,0),~(2\pi/3,-2\pi/\sqrt 3),~(2\pi/3,2\pi/\sqrt 3)$ and $K'$ points at $(4\pi/3,0),~(-2\pi/3,2\pi/\sqrt 3),~(-2\pi/3,-2\pi/\sqrt 3)$. Upon the action of strain, the lattice vectors change according to Eq(\ref{eq:DistanceCorrection}). The reciprocal lattice vectors, to leading order in strain, can be shown to change according to 
\beq\label{eq:A1}K_{X,\beta}=K^0_{X,\alpha}(\delta_{\alpha\beta}-u_{\alpha\beta})+\mathcal{O}(u_{\alpha\beta}^2),\eeq 
where $X\in\{\Gamma, M, K, K'\}$ as the case may be. In Sec. \ref{Subsec:UCon} we saw that $-K^0_{X,\alpha}u_{\alpha\beta}$ was  a part of the shift of the singular point. We now see that this part can be attributed to the shift of the BZ vector.

\section{$\bq$ and strain expansions for 1 atom per unit cell structures}\label{AppA}
\paragraph*{Square Lattice:} The Hamiltonian of a one-atom per cell square lattice within nn approximation is:
\begin{equation}\label{eq:SQHam}
    h^{\rm Sq}(\bk) = -2 t(\cos k_x + \cos k_y).
\end{equation}
At the $\Gamma$- and $M$-points of the BZ, the $\bq$-expansions to the Hamiltonian are:
\bea
    h^{\rm Sq}(\bK_{\Gamma}+\bq)&=&-4t+t(q_x^2+q_y^2)+\mathcal{O}(q^4),\nn\\
    h^{\rm Sq}(\bK_{M}+\bq)&=&~~4t-t(q_x^2+q_y^2)+\mathcal{O}(q^4).
\eea
The bond-wavevector group is $C_{4v}$ where the scalars are from the $A_1$ irrep with the form $1, x^2+y^2, (x^2+y^2)^2,$ etc. This is why the next leading order is $q^4$ and not $q^3$. The Hamiltonian expanded to $q^2$ readily allows us to extract the velocity and the inverse mass tensor. The leading order strain corrections to those points computed explicitly from Eq. (\ref{eq:strainExpansion}) are
\bea
    s^{\rm Sq}(\bK_{\Gamma}+\bq)&=&~~2t\zeta (\epsilon_{xx}+\epsilon_{yy}),\nn\\
    s^{\rm Sq}(\bK_{M}+\bq)&=&-2t\zeta (\epsilon_{xx}+\epsilon_{yy}).
\eea
\paragraph*{Triangular Lattice:} The one-atom per cell nn Hamiltonian is:
\begin{equation}\label{eq:THam0}
    h^{\rm Tr}(\bk) = 2 t -4 t\cos\frac{k_x}{2}\left(\cos\frac{k_x}{2}+\cos\frac{\sqrt3 k_y}{2}\right).
\end{equation}
At the $\Gamma$- and $K/K'$-points of the BZ, the $\bq$-expansions to the Hamiltonian are:
\bea
    h^{\rm Tr}(\bK_{\Gamma}+\bq) &=&-6t+\frac{3}{2}t(q_x^2+q_y^2)+\mathcal{O}(q^4),\nn\\
    h^{\rm Tr}(\bK_{K/K'}+\bq) &=&~~3t -\frac{3}{4}t(q_x^2+q_y^2)+\mathcal{O}(q^3).
\eea
The bond-wavevector group at $\Gamma$ is $C_{6v}$ where the scalars are from the $A_1$ irrep with the form $1, x^2+y^2, (x^2+y^2)^2,$ etc. This is why the next leading order is $q^4$.  The bond-wavevector group at $K,K'$ is $C_{3v}$ where the scalars are also from the $A_1$ irrep with the form $1, x^2+y^2, x(x^2-3y^2)$. This is why the next leading order is $q^3$ and not $q^4$ like the $\Gamma$-point. The leading order strain corrections, including both $s^D$ and $s^H$ corrections, to those points computed explicitly from Eq. (\ref{eq:strainExpansion}) are
\bea
    s^{\rm Tr}(\bK_{\Gamma}) &=&~~~3t\zeta (\epsilon_{xx}+\epsilon_{yy}),\nn\\
    s^{\rm Tr}(\bK_{K/K'}) &=&-\frac{3}{2}t\zeta (\epsilon_{xx}+\epsilon_{yy}).
\eea

\section{$\bq$ and strain expansions for 2 atom per unit cell structures}\label{AppB}
\paragraph*{Square Lattice:} We consider the interesting case of $\pi$-flux square lattice whose Hamiltonian is:
\begin{equation}\label{eq:PiHam}
    h^\pi_{ab}(\bk) = \begin{pmatrix} 0 & f(\bk)\\f^*(\bk)&0 \end{pmatrix},
\end{equation}
where
$$f(\bk)=-2t\left(\cos\frac{k_x+k_y}{2}-i\sin\frac{k_x-k_y}{2}\right).$$
At the $\Gamma$, $M$, Dirac (D) [at $\pm\frac\pi 2(1,1)$], and Parabolic (P) points [at $\pm \frac \pi 2(1,-1)$], the $\bq$-expansion of $f$ yields:
\bea\label{eq:qpi}
    f(\bK_{\Gamma}+\bq) &=& -t\left[2-i(q_x-q_y) -\frac14(q_x+q_y)^2\right] +\mathcal{O}(q^3),\nn\\
    f(\bK_{M}+\bq) &=& -t\left[-2-i(q_x-q_y) +\frac14(q_x+q_y)^2\right] +\mathcal{O}(q^3),\nn\\
    f(\bK_{D_\pm}+\bq) &=& \pm t\left[q_x+q_y\pm i(q_x-q_y)\right] +\mathcal{O}(q^3),\nn\\
    &=&\pm\sqrt{2}te^{\pm i\pi/4}(q_x\mp iq_y)+\mathcal{O}(q^3),\nn\\
    f(\bK_{P_\pm}+\bq) &=& -t\left[\left\{2-\frac14(q_x+q_y)^2\right\}\mp i\left\{2-\frac14(q_x-q_y)^2\right\}\right] +\mathcal{O}(q^4),\nn\\
    &=&\frac{t}{2\sqrt{2}}e^{\mp i\pi/4}\left[-8+q_x^2+q_y^2\pm2iq_xq_y\right] +\mathcal{O}(q^4).
\eea
This example is non-standard as it involves the presence of a magnetic flux. The symmetry group would involve magnetic translations and systems with $2\pi n/m$ flux linked with the original square plaquette, with $m$ being even, would have $m$ Dirac points. This enforcement arises from special gauge symmetries\cite{Wen1989} addressing which are beyond the scope of this work. Nevertheless, we can model the system in a particular gauge as a 2 atom per unit cell structure with only mirror as the symmetry. This structure belongs to the $C_{1v}$ group which allows for linear and quadratic terms to belong to the same scalar irrep $A_1$. The presence of the Dirac and parabolic points cannot be foreseen from spatial symmetries due to the presence of the magnetic flux, and arise from different considerations. But they are such that at the Dirac points the quadratic forms are absent while at the Parabolic points the linear terms are absent. 

The explicit calculation of the leading order strain corrections to these terms [leading to $f(\bk)+s(\bk)$ in the Hamiltonian] leads to
\bea\label{eq:spi}
    s(\bK_{\Gamma}) &=& t\zeta\left[\epsilon_{xx}+\epsilon_{yy}+2\epsilon_{xy}\right],\nn\\
    s(\bK_{M}) &=& it\left[e\mathcal A^M_x-e\mathcal A^M_y\right] -t\zeta\left[\epsilon_{xx}+\epsilon_{yy}+2\epsilon_{xy}\right],\nn\\
    s(\bK_{D_\pm}) &=& \pm\sqrt{2}te^{\pm i\pi/4}(e\mathcal A^{D_\pm}_x\mp ie\mathcal A^{D_\pm}_y),\nn\\
    s(\bK_{P_\pm}) &=& \sqrt2t\zeta e^{\mp i\pi/4}\left[\epsilon_{xx}+\epsilon_{yy}\pm2i\epsilon_{xy}\right],
\eea
where $e\mathcal A^X_\alpha=K_{X,\beta} u_{\beta\alpha}$, with $X\in\{M,D_+,D_-\}$.

\paragraph*{Triangular Lattice:} We consider the case of nn Honeycomb lattice whose Hamiltonian is:
\bea\label{eq:Grnn}
h^{\rm Gr}_{ab}(\bk) &=& \begin{pmatrix} 0& f(\bk)\\f^*(\bk)&0\end{pmatrix},\nn\\
\text{where, }f(\bk)&=&-t e^{-i\frac{k_y}{\sqrt3}}[1+2e^{i\frac{\sqrt3 k_y}{2}}\cos\frac{k_x}{2}].
\eea
The $\Gamma$- and $K$-point $\bq$ expansions are given by:
\bea\label{eq:qGr}
f(\bK_{\Gamma}+\bq)&=&-3t+\frac t4(q_x^2+q_y^2)+\mathcal{O}(q^3),\nn\\
f(\bK_{K}+\bq)&=&-\frac{\sqrt{3}t}{2}\left[q_x-iq_y\right] -\frac t8[q^2_x-q_y^2+2iq_xq_y] +\mathcal{O}(q^3).
\eea
The bond-wavevector group at $\Gamma$ is $C_{3v}$. The scalars can be constructed from the $A_1$ irrep with the form $1, x^2+y^2, x(x^2-3y^2)$, or from other irreps if we can find generators to support those irreps. This also explains why the next leading order is $q^3$. Since at $\Gamma$-point we already have $\hat h^0(\bK_\Gamma)=-3t\hat\sigma_x$ (with the $x$ Pauli matrix), $\hat \sigma_x$ has to belong to the $A_1$ irrep at $\Gamma$-point. We can also think of creating scalars using $x,y$ (to have finite velocity terms) that belong to $E$ irrep, but we would have to find a pair of generators that could belong to $E$ irrep. Since graphene is a $2\times2$ bipartite system, we can only have off-diagonal entries in the generator. This can only be possible by the pair $\{\sigma_x,\sigma_y\}$, but $\sigma_x$ has already been used to represent $A_1$. Thus, at $\Gamma$ we cannot have any Hilbert space generators for the vector irrep and hence there cannot be a velocity term. 

The bond-wavevector group at $K,K'$ is $C_{3v}$ as well. The scalars can again be from the $A_1$ irrep with the form $1, x^2+y^2, x(x^2-3y^2)$, but also from $E$ with form $\{x,y\}, \{x^2-y^2,xy\}, (x^2+y^2)\{x,y\},$ etc., if we can find generators in $E$. At this $\bK$ point, since $\hat h^0(\bK_{K/K'})=0$, we have the pair $\{\sigma_x,\sigma_y\}$ available and hence can represent the $E$ irrep. And precisely because of the bipartite nature and that $\hat\sigma_x,\hat\sigma_y$ have been used up for the E irrep, we don't have any generators left for the $A_1$ irrep, and thus there can be no $x^2+y^2$ term in the $\bq$ expansion. 

The explicit calculations of the leading order strain corrections to these terms [leading to $f(\bk)+s(\bk)$ in the Hamiltonian] are
\bea\label{eq:sGr}
s(\bK_{\Gamma})&=&\frac{3}{2}t\zeta(\epsilon_{xx}+\epsilon_{yy}),\nn\\
s(\bK_{K})&=&-\frac{\sqrt{3}t}{2}\left[e\mathcal A^K_x-ie\mathcal A^K_y\right]-\frac{3}{4}t\zeta[\epsilon_{xx}-\epsilon_{yy}+2i\epsilon_{xy}],
\eea
where $e\mathcal A^K_\alpha=K_{K,\beta} u_{\beta\alpha}$. The result at the $K'$ point is obtained by setting $\mathcal A_\alpha^{K'}=-\mathcal A_\alpha^{K'}$ and $s^H(-\bK)=s^{H*}(\bK)$. The latter results in $i\rightarrow -i$ in the $\e_{\alpha\beta}$ terms. If we were to include the nnn hopping ($t'$) of the honeycomb lattice, this would add a diagonal term to the Hamiltonian $h^{\rm Gr}_{ab}$ of the form $\delta_{ab}\bar f(\bk)$ where
$$\bar f(\bk)=2 t' -4 t'\cos\frac{k_x}{2}\left(\cos\frac{k_x}{2}+\cos\frac{\sqrt3 k_y}{2}\right).$$
The $\Gamma$- and $K/K'$-point $\bq$ expansions of this term are given by:
\bea\label{eq:qGrnn}
\bar f(\bK_{\Gamma}+\bq)&=&-6t'+\frac{3}{2}t'(q_x^2+q_y^2)+\mathcal{O}(q^4),\nn\\
\bar f(\bK_{K/K'}+\bq)&=&~~3t' -\frac{3}{4}t'(q_x^2+q_y^2)+\mathcal{O}(q^3).
\eea
This just acts like two copies of the 1 atom per unit cell triangular lattice case and the same results follow. The leading order strain corrections to this term [leading to $\bar f(\bk)+\bar s(\bk)$ in the Hamiltonian] are
\bea\label{eq:sGrnn}
    \bar s^{\rm Gr}(\bK_{\Gamma}) &=&~~~3t'\zeta (\epsilon_{xx}+\epsilon_{yy}),\nn\\
    \bar s^{\rm Gr}(\bK_{K/K'}) &=&-\frac{3}{2}t'\zeta (\epsilon_{xx}+\epsilon_{yy}).
\eea
Since $t'$ is in the diagonal, the scalar-potential in this case can be seen as a correction to the local chemical potential. But since the $s^H$ correction is entirely real, the correction is the same at the $K,K'$ points. This allows for the scalar-potential correction from nnn to be interpreted as a true chemical potential correction.

\section{$\bq$ and strain expansions for 3 atom per unit cell structures}\label{AppC}
\paragraph*{Square Lattice:}  Consider the Lieb lattice whose Hamiltonian is:
\begin{equation}\label{eq:LHam}
    h^{\rm Li}_{ab}(\bk) = -2t\begin{pmatrix} 0 & \cos\frac{k_x}2&\cos\frac{k_y}2\\\cos\frac{k_x}2&0&0\\\cos\frac{k_y}2&0&0\end{pmatrix}.
\end{equation}
The $\Gamma$- and $M$-point $\bq$ expansions are given by:
\bea\label{eq:qL}
 h^{\rm Li}_{ab}(\bK_\Gamma+\bq) &=& -2t\begin{pmatrix} 0 & 1-\frac{q_x^2}{8}&1-\frac{q_y^2}{8}\\
1-\frac{q_x^2}{8}&0&0\\
1-\frac{q_y^2}{8}&0&0\end{pmatrix}+\mathcal{O}(q^4),\nn\\
h^{\rm Li}_{ab}(\bK_M+\bq) &=& t\begin{pmatrix} 0 & q_x&q_y\\
q_x&0&0\\
q_y&0&0\end{pmatrix}+\mathcal{O}(q^3).
\eea
Here, the $\Gamma$ and $M$ points have $C_{4v}$ symmetry, but the bonds are $C_{2v}$. This leads to the bond-wavevector group to be $C_{2v}$. Each bond will form its own bond-wavevector group. Thus, $h_{12}(\bK)$ and $h_{13}(\bK)$ form two independent $C_{2v}$ bond-wavevector groups. Note that in the previous examples, the same bond contributed to all entries in the Hilbert space. But in systems with 3 or more atoms per unit cell, this is not always the case. Once again, the scalars can be formed from $A_1$ which contains $1,x^2,y^2$ functions. Since $\hat h^0(\bK_{\Gamma})\neq0$, the matrix structure at this point must belong to $A_1$. This matrix covers the entries corresponding to both bonds and hence no other entries can exist. This means that the matrix form of $\hat h^0(\bK_{\Gamma})$ used up the generator space to represent scalars leaving none to represent the vector irreps (which belong to $B_1$ and $B_2$). Thus, we cannot have any velocity elements at the $\Gamma$-point. The prevention of the mixing of the linear terms $x,y$ with the quadratic ones is the reason why the next contribution is $\mathcal{O}(q^4)$.

At the $M$-point, where the bond-wavevector group is still $C_{2v}$, $\hat h^0(\bK_M)=0$. This means that we can form matrices in the $B_1$ and $B_2$ irreps to contract with the vector forms $x,y$ in this group. However, in this case then, we cannot have any scalar forms $1,x^2,y^2$. This is why the next order contribution here is $\mathcal{O}(q^3)$.

Further, because of how the linear and quadratic forms are split, it is clear that $\Gamma$-point only has $s^H$ type corrections, whereas $M$-point has only $s^D$ type corrections. Indeed, an explicit calculation gives:
\bea\label{eq:sL}
    s^{\rm Li}_{ab}(\bK_{\Gamma}) &=&2t\zeta\begin{pmatrix} 0 & \epsilon_{xx}&\epsilon_{yy}\\
\epsilon_{xx}&0&0\\
\epsilon_{yy}&0&0\end{pmatrix},\nn\\
    s^{\rm Li}_{ab}(\bK_{M}) &=&t\begin{pmatrix} 0 & e\mathcal A^M_x&e\mathcal A^M_y\\
e\mathcal A^M_x&0&0\\
e\mathcal A^M_y&0&0\end{pmatrix},
\eea
where $e\mathcal A^M_\alpha=K_{M,\beta} u_{\beta\alpha}$.

\paragraph*{Triangular Lattice:} Consider the Kagom\'{e} lattice whose Hamiltonian is:
\begin{equation}\label{eq:KHam}
    h^{\rm Kag}_{ab}(\bk) = -2t\begin{pmatrix} 0 & \cos\frac{k_x+\sqrt{3}k_y}{4}&\cos\frac{k_x}{2}\\\cos\frac{k_x+\sqrt{3}k_y}{4}&0&\cos\frac{k_x-\sqrt{3}k_y}{4}\\\cos\frac{k_x}{2}&\cos\frac{k_x-\sqrt{3}k_y}{4}&0\end{pmatrix}.
\end{equation}
The $\Gamma$- and $K$-point $\bq$ expansions are given by:
\bea\label{eq:qK}
 h^{\rm Kag}_{ab}(\bK_\Gamma+\bq) &=& -2t\begin{pmatrix} 0 & 1-\frac{(q_x+\sqrt{3}q_y)^2}{32}&1-\frac{q_x^2}{8}\\
1-\frac{(q_x+\sqrt{3}q_y)^2}{32}&0&1-\frac{(q_x-\sqrt{3}q_y)^2}{32}\\
1-\frac{q_x^2}{8}&1-\frac{(q_x-\sqrt{3}q_y)^2}{32}&0\end{pmatrix}+\mathcal{O}(q^4),\nn\\
h^{\rm Kag}_{ab}(\bK_K+\bq) &=& t\begin{pmatrix} 
0 &-1-\frac{\sqrt{3}q_x+3q_y}{4}+\frac{(q_x+\sqrt{3}q_y)^2}{32}&1-\frac{\sqrt{3}q_x}{2}-\frac{q_x^2}{8}\\
-1-\frac{\sqrt{3}q_x+3q_y}{4}+\frac{(q_x+\sqrt{3}q_y)^2}{32}&0&-1-\frac{\sqrt{3}q_x-3q_y}{4}+\frac{(q_x-\sqrt{3}q_y)^2}{32}\\
1-\frac{\sqrt{3}q_x}{2}-\frac{q_x^2}{8}&-1-\frac{\sqrt{3}q_x-3q_y}{4}+\frac{(q_x-\sqrt{3}q_y)^2}{32}&0
\end{pmatrix}+\mathcal{O}(q^3).\nn\\
\eea
Here the bonds are also $C_{2v}$ and bond-wavevector group is $C_{2v}$ at the $\Gamma$-point, while at $K,K'$ they are $C_{1v}$ (the common subgroup of $C_{3v}$ and $C_{2v}$). At the $\Gamma$-point the analysis is the same as above and we cannot have the linear functions $x,y$ due to $\hat h^0(\bK_\Gamma)\neq0$. The next order also starts from $q^4$. Since the group is $C_{2v}$, the $x,y$ or $x^2,y^2$ forms correspond to the distances along the direction of the bond in whatever co-ordinate system. This is the reason for the grouping of $q_x,q_y$ in Eq. (\ref{eq:qK}): those correspond to $\bq\cdot\boldsymbol{\delta}_{ab}$.

At the $K,K'$ points however, both the linear and quadratic functions are allowed as the group is now $C_{1v}$. Once again the $x,x^2$ are to be replaced by the distances along the bonds. The leading order strain corrections also follow this picture leading to:
\bea\label{eq:sK}
 s^{\rm Kag}_{ab}(\bK_\Gamma) &=& t\zeta\begin{pmatrix} 0 &\frac{\epsilon_{xx}+3\epsilon_{yy}+2\sqrt{3}\epsilon_{xy}}{2}&2\epsilon_{xx}\\
\frac{\epsilon_{xx}+3\epsilon_{yy}+2\sqrt{3}\epsilon_{xy}}{2}&0&\frac{\epsilon_{xx}+3\epsilon_{yy}-2\sqrt{3}\epsilon_{xy}}{2}\\
2\epsilon_{xx}&\frac{\epsilon_{xx}+3\epsilon_{yy}-2\sqrt{3}\epsilon_{xy}}2&0
\end{pmatrix},\nn\\
s^{\rm Kag}_{ab}(\bK_K) &=& t\begin{pmatrix} 
0 &-\frac{\sqrt{3}e\mathcal A^K_x+3e\mathcal A^K_y}{4}+\zeta\frac{\epsilon_{xx}+3\epsilon_{yy}+2\sqrt{3}\epsilon_{xy}}{4}&-\frac{\sqrt{3}e\mathcal A^K_x}{2}-\zeta\epsilon_{xx}\\
-\frac{\sqrt{3}e\mathcal A^K_x+3e\mathcal A^K_y}{4}+\zeta\frac{\epsilon_{xx}+3\epsilon_{yy}+2\sqrt{3}\epsilon_{xy}}{4}&0&-\frac{\sqrt{3}e\mathcal A^K_x-3e\mathcal A^K_y}{4}+\zeta\frac{\epsilon_{xx}+3\epsilon_{yy}-2\sqrt{3}\epsilon_{xy}}{4}\\
-\frac{\sqrt{3}e\mathcal A^K_x}{2}-\zeta\epsilon_{xx}&-\frac{\sqrt{3}e\mathcal A^K_x-3e\mathcal A^K_y}{4}+\zeta\frac{\epsilon_{xx}+3\epsilon_{yy}-2\sqrt{3}\epsilon_{xy}}{4}&0
\end{pmatrix},\nn\\
\eea
where $e\mathcal A^K_\alpha=K_{K,\beta}u_{\beta\alpha}$. The Hamiltonian at $K'$ point is obtained in the same way as was done in the triangular lattice section in Appendix \ref{AppB}.

\section{$\bq$ and strain expansions for a large Hilbert space}\label{AppD}
Consider the super-Honeycomb lattice with 5 atoms per unit cell. The Hamiltonian is:
\bea\label{eq:HKHam}
h_{ab}^{\rm sH}(\bk)&=& \begin{pmatrix}
        0_{2\times2} & \hat b_{2\times3}(\bk)\\
        \hat b^\dag_{3\times2}(\bk)&0_{3\times3}
    \end{pmatrix},\nn\\
\text{where, }\hat b(\bk)&=&-t\begin{pmatrix}
       e^{-i\frac{k_y+\sqrt3 k_x}{4\sqrt3}}& e^{i\frac{k_y}{2\sqrt3}} & e^{-i\frac{k_y-\sqrt3 k_x}{4\sqrt3}}\\
        e^{i\frac{k_y+\sqrt3 k_x}{4\sqrt3}}& e^{-i\frac{k_y}{2\sqrt3}} & e^{i\frac{k_y-\sqrt3 k_x}{4\sqrt3}}
    \end{pmatrix}.
\eea
The $\Gamma$- and $K$-point $\bq$ expansions are given by:
\bea\label{eq:qHK}
 \hat b(\bK_\Gamma+\bq)&=&-t\begin{pmatrix}
       1-ir_1-r_1^2/2& 1+ir_2-r_2^2/2 & 1-ir_3-r_3^2/2\\
       1+ir_1-r_1^2/2& 1-ir_2-r_2^2/2 & 1+ir_3-r_3^2/2\\
    \end{pmatrix}+\mathcal{O}(q^3),\nn\\
    \hat b(\bK_K+\bq)&=&-t\begin{pmatrix}
       e^{i\pi/3}(1-ir_1-r_1^2/2)& 1+ir_2-r_2^2/2 & e^{-i\pi/3}(1-ir_3-r_3^2/2)\\
       e^{-i\pi/3}(1+ir_1-r_1^2/2)& 1-ir_2-r_2^2/2 & e^{i\pi/3}(1+ir_3-r_3^2/2)\\
    \end{pmatrix}+\mathcal{O}(q^3).
\eea
where $$r_1\equiv\frac{q_y+\sqrt3 q_x}{4\sqrt3};~r_2\equiv\frac{q_y}{2\sqrt3};~r_3\equiv\frac{q_y-\sqrt3 q_x}{4\sqrt3}.$$
In this case the groups everywhere are $C_{1v}$ and thus fall under the same category as the Kagom\'e lattice. The leading order strain corrections (which result in $\hat b+\hat s$ in the Hamiltonian) are
\bea\label{eq:sHK}
\hat s(\bK_\Gamma)&=&\frac14t\zeta\begin{pmatrix}
       3\epsilon_{xx}+\epsilon_{yy}+2\sqrt{3}\epsilon_{xy}& 4\epsilon_{yy} & 3\epsilon_{xx}+\epsilon_{yy}-2\sqrt{3}\epsilon_{xy}\\
       3\epsilon_{xx}+\epsilon_{yy}+2\sqrt{3}\epsilon_{xy}& 4\epsilon_{yy} & 3\epsilon_{xx}+\epsilon_{yy}-2\sqrt{3}\epsilon_{xy}\\
    \end{pmatrix},\nn\\
\hat s(\bK_K)&=&\frac14t\zeta\begin{pmatrix}
       e^{i\pi/3}[3\epsilon_{xx}+\epsilon_{yy}+2\sqrt{3}\epsilon_{xy}]& 4\epsilon_{yy} & e^{-i\pi/3}[3\epsilon_{xx}+\epsilon_{yy}-2\sqrt{3}\epsilon_{xy}]\\
       e^{-i\pi/3}[3\epsilon_{xx}+\epsilon_{yy}+2\sqrt{3}\epsilon_{xy}]& 4\epsilon_{yy} & e^{i\pi/3}[3\epsilon_{xx}+\epsilon_{yy}-2\sqrt{3}\epsilon_{xy}]
    \end{pmatrix}\nn\\
    &&-t\begin{pmatrix}
       -ie^{i\pi/3}A_1& iA_2& -ie^{-i\pi/3}A_3\\
       ie^{-i\pi/3}A_1& -iA_2& ie^{i\pi/3}A_3\\
    \end{pmatrix},
\eea
where $A_i$ is the same as $r_i$ with $q_\alpha\rightarrow e\mathcal A^K_\alpha\equiv K_{K,\beta}u_{\beta\alpha}$.

\bibliography{References.bib}
\end{document}